\documentclass[nature,twocolumn,showkeys,superscriptaddress]{revtex4-1}

\usepackage{placeins}
\usepackage[artemisia]{textgreek}
\usepackage{amssymb}
\usepackage{epsfig}
\usepackage{amsmath}
\usepackage{graphicx}
\usepackage{dcolumn}
\usepackage{latexsym}
\usepackage{color}
\usepackage{subfigure}
\usepackage{nicefrac}
\usepackage{blindtext}
\usepackage[ulem=normalem]{changes}
\usepackage{float}
\usepackage{soul}
\usepackage[T1]{fontenc}
\usepackage[hidelinks]{hyperref}
\setcitestyle{super}

\hypersetup{
    colorlinks,
    linkcolor={red!50!black},
    citecolor={blue!50!black},
    urlcolor={blue!80!black}
}

\begin{document}

\title{Atomically sharp domain walls in an antiferromagnet}

\author{Filip Krizek}
\email{krizekfi@fzu.cz}
\affiliation{Institute of Physics, Czech Academy of Sciences, Cukrovarnick{\'a} 10, 162 00 Praha 6, Czech Republic.}

\author{Sonka Reimers}
\affiliation{School of Physics and Astronomy, University of Nottingham, Nottingham NG7 2RD, United Kingdom}
\affiliation{Diamond Light Source, Harwell Science and Innovation Campus, Didcot, Oxfordshire, OX11 ODE, United Kingdom}

\author{Zden\v{e}k Ka\v{s}par}
\affiliation{Institute of Physics, Czech Academy of Sciences,  Cukrovarnick{\'a} 10, 162 00 Praha 6, Czech Republic.}
\affiliation{Faculty of Mathematics and Physics, Charles University, Ke Karlovu 3, 121 16 Prague 2, Czech Republic}

\author{Alberto Marmodoro}
\affiliation{Institute of Physics, Czech Academy of Sciences,  Cukrovarnick{\'a} 10, 162 00 Praha 6, Czech Republic.}

\author{Jan Michali\v{c}ka}
\affiliation{Central European Institute of Technology, Brno University of Technology, Purky\v{n}ova 123, 612 00, Brno, Czech Republic}

\author{Ond\v{r}ej Man}
\affiliation{Central European Institute of Technology, Brno University of Technology, Purky\v{n}ova 123, 612 00, Brno, Czech Republic}

\author{Alexander Edstr\"{o}m}
\affiliation{Institut de Ciencia de Materials de Barcelona (ICMAB-CSIC), Campus UAB, 08193 Bellaterra, Spain}

\author{Oliver J. Amin}
\affiliation{School of Physics and Astronomy, University of Nottingham, Nottingham NG7 2RD, United Kingdom}

\author{Kevin W. Edmonds}
\affiliation{School of Physics and Astronomy, University of Nottingham, Nottingham NG7 2RD, United Kingdom}

\author{Richard P. Campion}
\affiliation{School of Physics and Astronomy, University of Nottingham, Nottingham NG7 2RD, United Kingdom}

\author{Francesco Maccherozzi}
\affiliation{Diamond Light Source, Harwell Science and Innovation Campus, Didcot, Oxfordshire, OX11 ODE, United Kingdom}

\author{Sarnjeet S. Dhesi}
\affiliation{Diamond Light Source, Harwell Science and Innovation Campus, Didcot, Oxfordshire, OX11 ODE, United Kingdom}

\author{Jan Zub{\'a}\v{c}}
\affiliation{Institute of Physics, Czech Academy of Sciences,  Cukrovarnick{\'a} 10, 162 00 Praha 6, Czech Republic.}
\affiliation{Faculty of Mathematics and Physics, Charles University, Ke Karlovu 3, 121 16 Prague 2, Czech Republic}

\author{Jakub~\v{Z}elezn\'y}
\affiliation{Institute of Physics, Czech Academy of Sciences,  Cukrovarnick{\'a} 10, 162 00 Praha 6, Czech Republic.}

\author{Karel~V\'yborn\'y}
\affiliation{Institute of Physics, Czech Academy of Sciences,  Cukrovarnick{\'a} 10, 162 00 Praha 6, Czech Republic.}

\author{Kamil Olejn{\'i}k}
\affiliation{Institute of Physics, Czech Academy of Sciences,  Cukrovarnick{\'a} 10, 162 00 Praha 6, Czech Republic.}

\author{V{\'i}t Nov{\'a}k}
\affiliation{Institute of Physics, Czech Academy of Sciences,  Cukrovarnick{\'a} 10, 162 00 Praha 6, Czech Republic.}

\author{J{\'a}n R{\'u}sz}
\affiliation{Department of Physics and Astronomy, Uppsala University, Box 516, 75120 Uppsala, Sweden}

\author{Juan Carlos Idrobo}
\affiliation{Center for Nanophase Materials Sciences, Oak Ridge National Laboratory, Oak Ridge, Tennessee 37831, USA}

\author{Peter Wadley}
\affiliation{School of Physics and Astronomy, University of Nottingham, Nottingham NG7 2RD, United Kingdom}

\author{Tomas Jungwirth}
\email{jungw@fzu.cz}
\affiliation{Institute of Physics, Czech Academy of Sciences,  Cukrovarnick{\'a} 10, 162 00 Praha 6, Czech Republic.}
\affiliation{School of Physics and Astronomy, University of Nottingham, Nottingham NG7 2RD, United Kingdom}

\date{\today}

\begin{abstract}
The interest in understanding scaling limits of magnetic textures such as domain walls spans the entire field of magnetism from its relativistic quantum fundamentals to applications in information technologies. The traditional focus of the field on ferromagnets has recently started to shift towards antiferromagnets which offer a rich materials landscape and utility in  ultra-fast and neuromorphic devices insensitive to magnetic field perturbations. Here we report the observation that domain walls in an epitaxial crystal of antiferromagnetic CuMnAs  can be atomically sharp. We reveal this ultimate domain wall scaling limit using differential phase contrast imaging within aberration-corrected scanning transmission electron microscopy, which we complement by X-ray magnetic dichroism microscopy and ab initio calculations. We highlight that the atomically sharp domain walls are outside the remits of established spin-Hamiltonian theories and can offer device functionalities unparalleled in ferromagnets.
\end{abstract}

\pacs{}

\maketitle

Magnetic textures such as domain walls or vortices provide a basic test-bed for our physical understanding of magnetic systems  \cite{Hellman2016}. From an applied perspective, when representing bits in information technologies, the texture dimensions determine fundamental scaling limits for the data density \cite{Parkin2015}.  Continuum micromagnetic theories (or their more elaborate atomistic variants) based on model spin Hamiltonians represent a powerful technique for predicting the morphology of non-uniform magnetic textures \cite{Hellman2016}. Among the spin interactions commonly considered in micromagnetics, the exchange energy prefers collinear alignment of neighboring moments, i.e. large spatial scales of the rotating spins in the texture, while the anisotropy energy favors more abrupt spin reorientations. 

In common bulk magnets such as Fe or Co, a significantly larger energetic contribution of exchange interaction with respect to anisotropy  results in typical domain wall widths that exceed interatomic distances by orders of magnitude. Nanometer-scale domain walls were observed in rare earth magnets with large magnetic anisotropies due to strongly relativistic heavy elements \cite{Lloyd2002}, or in fragile low-temperature  systems comprising a highly anisotropic mono-atomic layer of a magnet deposited on a heavy-element substrate \cite{Bode2006}. 

In this paper, we report an experimental observation of the ultimate limit  of the domain wall width, in which the reorientation of the magnetic order between the opposite domains occurs abruptly at neighboring atomic sites in the crystal. The choice of the system where we detect the atomically sharp domain walls has not been guided by micromagnetics but rather counters its expectations. We make the observation in an epitaxially grown 50~nm thick film of  the weakly anisotropic  antiferromagnet CuMnAs, which is composed of common light elements and has a strong exchange energy with a transition to the N\'eel magnetic order well above room temperature \cite{Wadley2016,Wang2020}. 

Indeed, earlier measurements in the CuMnAs antiferromagnet by X-ray magnetic linear dichroism photoemission electron microscopy (XMLD-PEEM) identified domain walls of a large $\approx100$~nm width, as expected from its weak anisotropy and  strong exchange coupling \cite{Wadley2018}. Here we will refer to these textures as  the micromagnetic domain walls. The experiments also demonstrated a reversible motion of the micromagnetic walls controlled by applied current pulses of opposite polarity. The electrical reorientation of the N\'eel vector  was achieved owing to the specific crystallographic structure and antiferromagnetic ordering of CuMnAs \cite{Zelezny2014,Wadley2018}. 

\begin{figure*}[htbp!]
\vspace{0.2cm}
\includegraphics[scale=1]{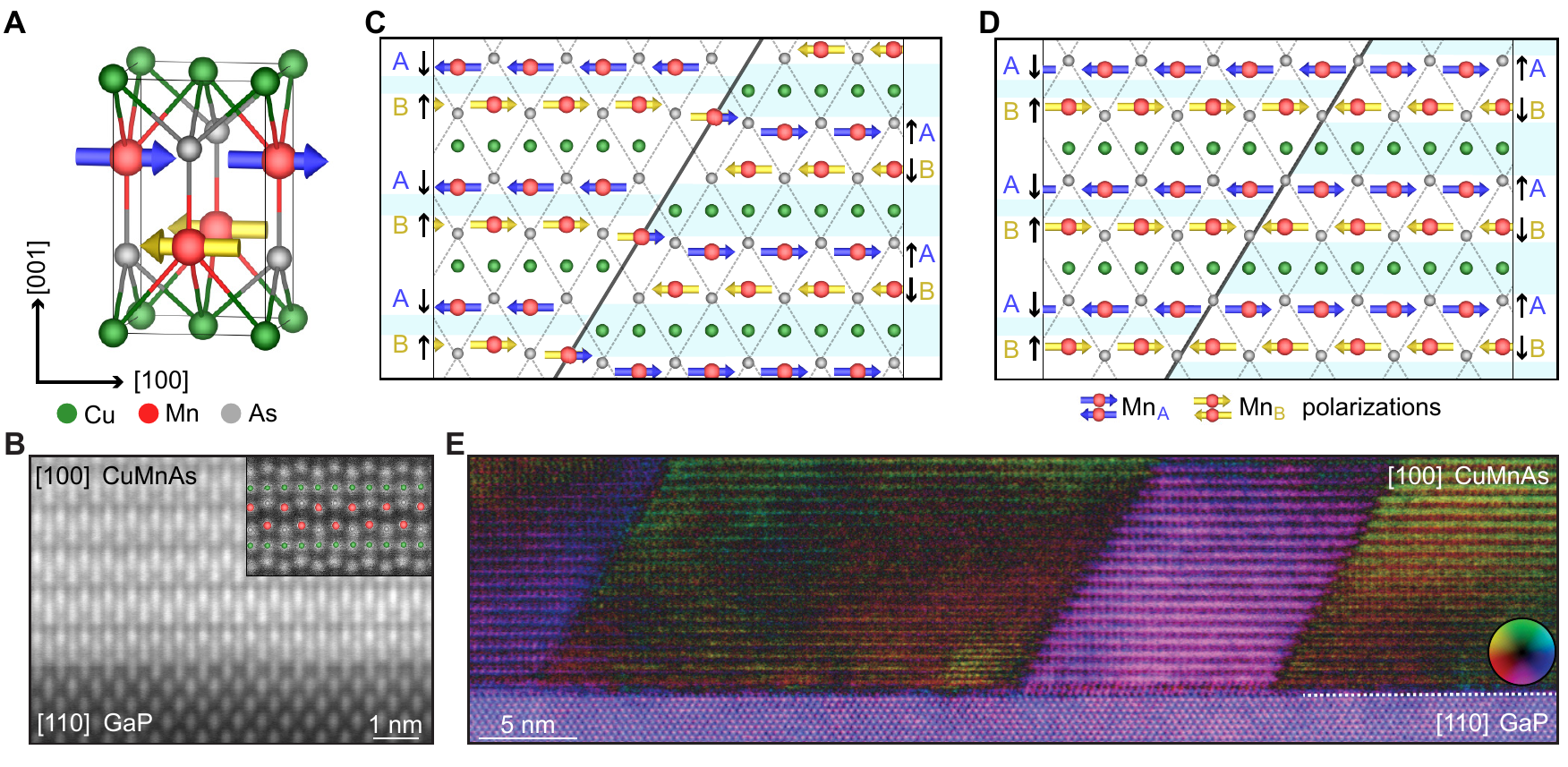}
\vspace{-0.4cm}
\caption{{\bf{Crystal structure and atomically sharp domain walls in antiferromagnetic CuMnAs.}} ({\bf A}) Atomic model of the CuMnAs unit-cell. ({\bf B}) HAADF-STEM image of a [100] projection of the epitaxial CuMnAs film grown on lattice-matched GaP. ({\bf C,D}) Schematics of the atomically sharp domain walls at an anti-phase boundary defect and in an unperturbed area of the CuMnAs single-crystal, respectively. Symbols A (blue) and B (yellow) label upper and lower Mn sublattices from the unit-cell in panel (A). Thin dashed lines highlight preserved As atom matrix. Black arrows represent Lorentz force direction at individual sublattices, which focuses the deflected beam into the areas with light blue overlay. ({\bf E})  An overview DPC-STEM image of the atomically sharp domain walls in a CuMnAs film.} 
\label{fig1}
\end{figure*}

In  Fig.~\ref{fig1}A we show a schematic of the CuMnAs unit cell and in Fig.~\ref{fig1}B an atomically resolved annular dark-field (Z-contrast) scanning transmission electron microscopy (STEM) image of the epitaxially grown CuMnAs film. The crystal has two non-centrosymmetric Mn sites per unit cell with antiparallel moments (schematically shown by the two color arrows in Fig.~\ref{fig1}A) that  are crystallographically distinct and thus not connected by a simple lattice translation. This special symmetry  of the Mn sites, which allowed for the electrical N\'eel vector reorientation, plays also an important role in facilitating our observation of the atomically sharp domain walls in the same antiferromagnetic material.

Prior to our present work, a hint towards new rich physics beyond the micromagnetics expectations has been recently shown in a study reporting  unconventional functionalities in CuMnAs memory devices \cite{Kaspar2019}. In the study, a distinct unipolar switching mechanism has been experimentally demonstrated, with excitation times scaled down to a single femtosecond-laser pulse, readout signals reaching giant-magnetoresistance amplitudes, and with analog neuromorphic-like switching and retention characteristics\cite{Kaspar2019}. The unconventional switching mechanism has been ascribed to quenching the antiferromagnet into nano-fragmented domain states with inferred scales of the textures below the $\sim 10$~nm resolution limits of the employed scanning NV-diamond magnetometry and XMLD-PEEM \cite{Wornle2019,Kaspar2019}. These scales are significantly smaller than the width of the micromagnetic domain walls in the material.

\begin{figure*}[htbp!]
\vspace{0.2cm}
\includegraphics[scale=1]{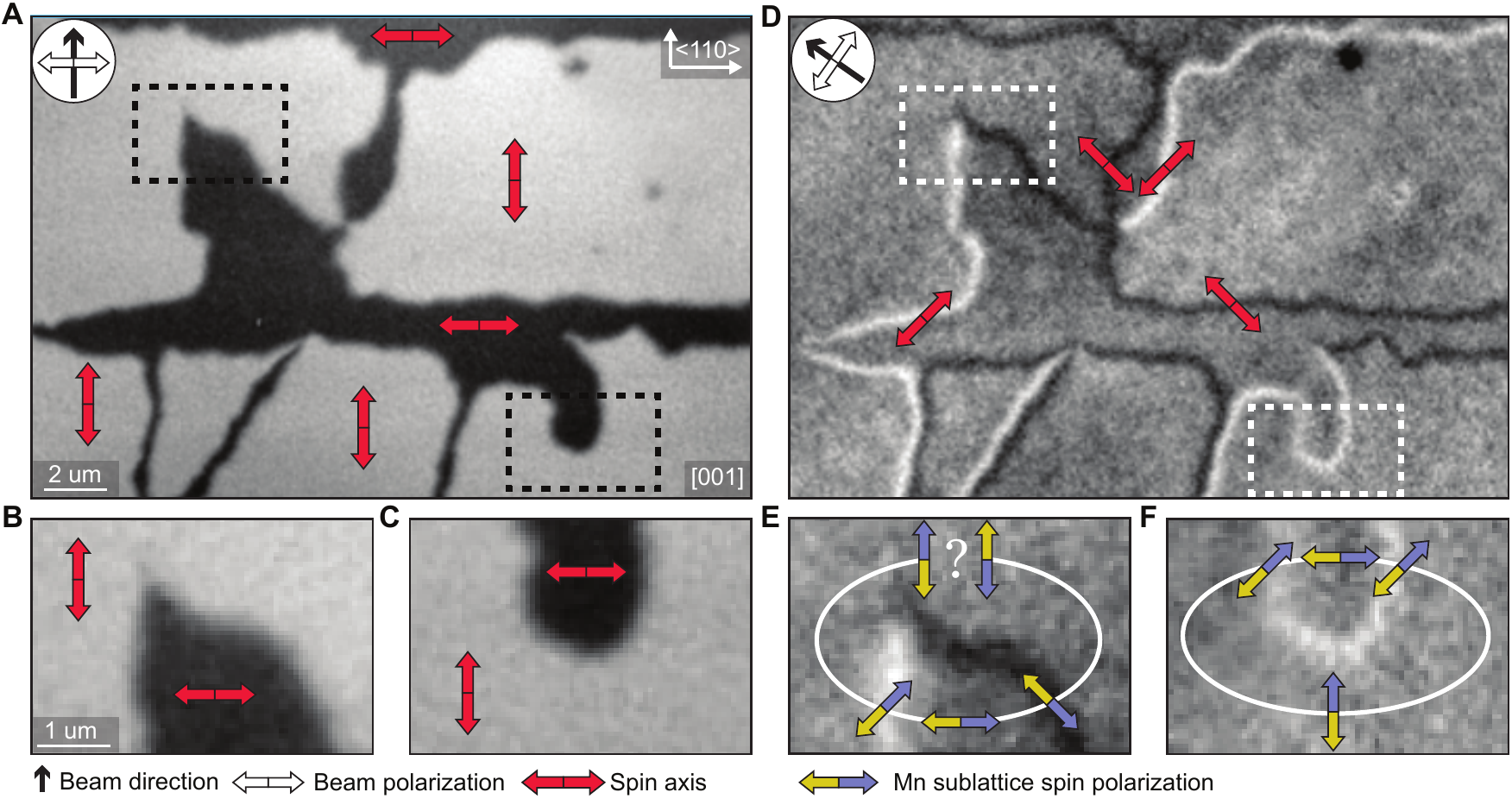}
\vspace{-0.4cm}
\caption{{\bf{The presence of sharp 180$^{\circ}$ domain walls inferred from XMLD-PEEM.}} ({\bf A}) XMLD-PEEM micrograph of the surface of the CuMnAs film. The compass indicates the direction of the X-ray beam and the white double arrow its polarization. Red double-arrows indicate the spin axis of selected antiferromagnetic domains corresponding to the measured black/white contrast. ({\bf B,C}) Zoom-ins on two regions selected from {({A})}. ({\bf D}) XMLD-PEEM micrograph corresponding to the area in (A) with the beam direction and polarization rotated by 45$^{\circ}$. Red double-arrows correspond to the mean angle of the spin axis in the micromagnetic domain walls. ({\bf E,F}) Zoom-ins on the same regions as in (B) where the blue and yellow arrows indicate Mn$_{\rm A}$ and Mn$_{\rm B}$ sublattice moments, respectively, i.e., the orientation of the N\'eel vector. The N\'eel vector returns to its original orientation when closing a loop  in (F). In contrast, the N\'eel vector appears to be reversed when completing the closed loop  in (E), indicating that a 180$^{\circ}$ reversal had to occur odd number of times along the loop and that the corresponding sharp domain wall(s) is below the XMLD-PEEM resolution.}
\label{fig2}
\end{figure*}

The atomically sharp domain walls in CuMnAs, discovered in the present work, are identified by the differential phase contrast (DPC) STEM imaging technique \cite{Dekker1974,Chapman1978,Muller2014,Shibata2015,Lohr2015,Lazic2016,Matsumoto2016,Yucelen2018,Chen2018c,Hachtel2018,Edstrom2019}.  DPC produces an image that reflects the relative shifts observed on the convergent beam electron diffraction (CBED) disks of an atomic size electron probe due to a material's local electric and magnetic fields \cite{Dekker1974,Chapman1978}. A lower-resolution overview image of the antiferromagnetic CuMnAs film illustrating the sharp changes in the DPC-STEM contrast  is shown in  Fig.~\ref{fig1}E. In the main body of the paper we will show that at high resolution we can associate these DPC-STEM signals with two types of abrupt  N\'eel vector reversals, schematically illustrated in Figs.~\ref{fig1}C and 1D:  The first type occurs at a crystallographic anti-phase boundary defect (Figs.~\ref{fig1}C), while the second type forms in an unperturbed part of the crystal (Figs.~\ref{fig1}D). 

The paper is organized as follows: We start from magnetic images of our films taken by the established, but low-resolution XMLD-PEEM technique, providing an indirect evidence for the presence of narrow  antiferromagnetic domain walls in our films. Next, we present the experimental and theoretical analysis of the high-resolution DPC-STEM images of the  atomically sharp domain walls, and discuss their {\em ab initio} modelling. Finally, before concluding the paper, we rule out structural artifact interpretations of the DPC-STEM domain wall images by systematically scrutinizing scenarios of abruptly varying  strain, chemical composition, lamella thickness, crystal rotation, and formation of crystal grain overlaps.

The XMLD-PEEM data revealing that 180$^\circ$ N\'eel vector reversals on scales below  $\sim 10$~nm are present in the material are shown in Fig.~\ref{fig2} (see Supplementary text for details on the XMLD-PEEM method). The evidence follows from tracking the N\'eel vector reorientations along closed paths encompassing well resolved biaxial domains and 90$^\circ$ micromagnetic domain walls in the epitaxial CuMnAs film grown on a lattice matched GaP substrate \cite{Wadley2018}. With the X-ray polarization along one of the $\langle 110\rangle$ magnetic easy axes of the biaxial CuMnAs, we observe a strong black and white contrast distinguishing micron-size domains with the N\'eel vector aligned with either the [110] or [1$\bar{1}$0] axis (Fig.~\ref{fig2}A-C). Here we recall that the XMLD-PEEM contrast can only resolve N\'eel vectors aligned along different axes while it is insensitive to the sign of the N\'eel vector.  To identify the rotation angle of the N\'eel vector axis in the domain walls separating the [110]/[1$\bar{1}$0] domains, we align the X-ray polarization along one of the $\langle 100\rangle$  directions (Fig.~\ref{fig2}D-F). In these measurements we image the 90$^\circ$  micromagnetic domain walls of $\approx 100$~nm width. The black and white contrast distinguishes between the mean axes of the N\'eel vector in the domain wall along either [100] or [010] directions.

The observed strong contrasts for both X-ray polarizations allow us to track not only the N\'eel vector axis but also the vector itself which makes a $\pm45^\circ$ rotation whenever crossing from a domain to a micromagnetic domain wall or {\em vice versa}.  For example, starting from an arbitrary but fixed definition of the sign of the N\'eel vector in the bottom middle part of Fig.~\ref{fig2}F, one can proceed along a closed loop intersecting two micromagnetic domain walls. Since in this case the two domain walls have the same (white) contrast, the  N\'eel vector returns to its original direction and sign when completing the loop. Remarkably, in Fig.~\ref{fig2}E, the closed loop intersects two micromagnetic domain walls of opposite contrast. This implies that the N\'eel vector flipped sign an odd-number of times and that the length-scale of the underlying sharp domain wall (walls) is below the  $\sim 10$~nm resolution of XMLD-PEEM.

\begin{figure*}[htbp!]
\vspace{0.2cm}
\includegraphics[scale=1]{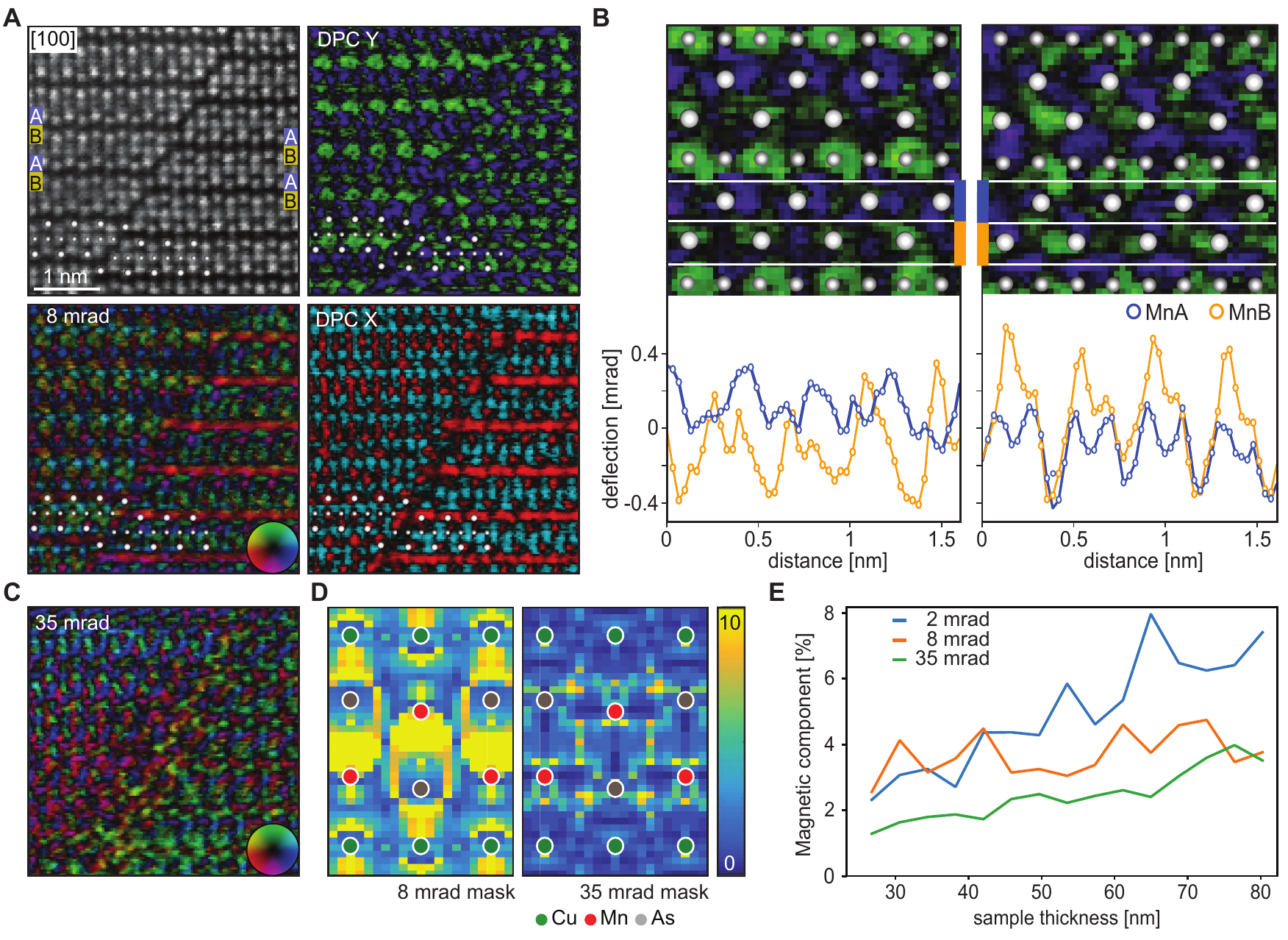}
\vspace{-0.4cm}
\caption{{\bf{DPC-STEM measurements of atomically sharp domain walls at an anti-phase boundary.}} ({\bf A}) HAADF micrograph of an anti-phase boundary defect is shown in the left top panel, overlaid with a model of the CuMnAs crystal (for clarity we show only Mn and Cu atoms). Symbols A (blue) and B (yellow) label the upper and lower Mn sublattices. DPC-STEM image of a corresponding area, showing the total (color-circle symbol), the up/down (DPC-Y) and the left/right (DPC-X) deflection of the beam calculated from the center of mass shift of the ronchigram for each pixel of the HAADF-STEM image. The applied ronchigram mask radius is 8 mrad.  ({\bf B}) Zoom-ins of the DPC-Y image in (A) from areas left and right from the anti-phase boundary. The horizontal line profiles below the image show deflection averaged separately over the Mn$_{\rm A}$ (blue) and Mn$_{\rm B}$ (yellow) sublattice. ({\bf C}) Same total DPC-STEM image as in (A) processed with a 35 mrad ronchigram mask. ({\bf D}) Simulation of the distribution of the relative strength of the magnetic component with respect to the total DPC-STEM signal. The color scale is limited to 10\%. The atomic positions are highlighted by green/Cu, red/Mn, grey/As. ({\bf E}) Simulated dependence of the relative strength of the magnetic component of the DPC-STEM signal on sample thickness for three different sizes of the ronchigram mask.}
\label{fig3}
\end{figure*}

The antiferromagnetically aligned moments sitting on two distinct Mn crystal sublattices, not connected by a spatial translation, make CuMnAs a particularly favorable material for identifying  the atomically sharp antiferromagnetic domain walls by DPC-STEM. First we focus on the  domain wall located on the anti-phase boundary defect (see Fig.~\ref{fig1}C). This crystallographic defect, identified in earlier structural STEM measurements \cite{Krizek2020}, has a form of a lattice slip-dislocation propagating along $\{$011$\}$ planes and is a  consequence of the epitaxial growth of CuMnAs on the III-V substrate. The tetragonal CuMnAs lattice may start bonding to the substrate either with the lower or upper As plane (see Fig.~\ref{fig1}A),  corresponding to a change in the initial stacking of the Mn planes in the individual grains. As a result, $\approx c/3$ lattice-shift anti-phase boundaries form when islands with different stacking coalesce during further growth. At the anti-phase boundary, one of the Mn crystal sublattices (say Mn$_{\rm A}$) closely aligns with the other Mn crystal sublattice (Mn$_{\rm B}$). The anti-phase boundary, therefore, acts as a source for the formation of an atomically sharp magnetic domain wall with Mn$_{{\rm A}\rightarrow}$ and Mn$_{{\rm B}\leftarrow}$ on one side of the boundary and Mn$_{{\rm A}\leftarrow}$ and Mn$_{{\rm B}\rightarrow}$ on the other side (see Fig.~\ref{fig1}C). Simultaneously, the unperturbed non-magnetic crystal structures on either side of the  anti-phase boundary are indistinguishable, which makes the boundary an ideal object  for detecting the magnetic configuration by DPC-STEM. The whole crystal structure is fixed by the matrix of As/P atoms, which extends from the GaP substrate to the CuMnAs layer \cite{Krizek2020}, and is also preserved over the anti-phase boundary and defines its angle, as illustrated in Figs. \ref{fig1}B,C.

Figure~\ref{fig3}A (upper left panel) shows a high-angle annular dark-field (HAADF) STEM image of an anti-phase boundary in CuMnAs and the simultaneously acquired high-resolution DPC-STEM image (lower left panel). The latter was reconstructed by calculating the shifts of the center of mass of the recorded CBED patterns, known also as ronchigrams, for each pixel of the HAADF-STEM  image (see the Supplementary Information for experimental details). The right panels of Fig. \ref{fig3}A show the separated X and Y components of the calculated  shifts of the ronchigram's center of mass, corresponding to the [010] and [001] crystallographic axes of CuMnAs. In all three DPC-STEM images, there is a clear difference in contrast and intensity of the signal between the two sides  of the anti-phase boundary. Figure~\ref{fig3}B shows a zoom-in on a portion of the Y-component of the center of mass  signal taken from an area with $4\times4$ Mn atoms. We single out the center of mass Y-component here because the associated Lorentz force due the magnetic deflection of the Mn moments is vertically aligned. 

\begin{figure*}[htbp!]
\vspace{0.2cm}
\includegraphics[scale=1]{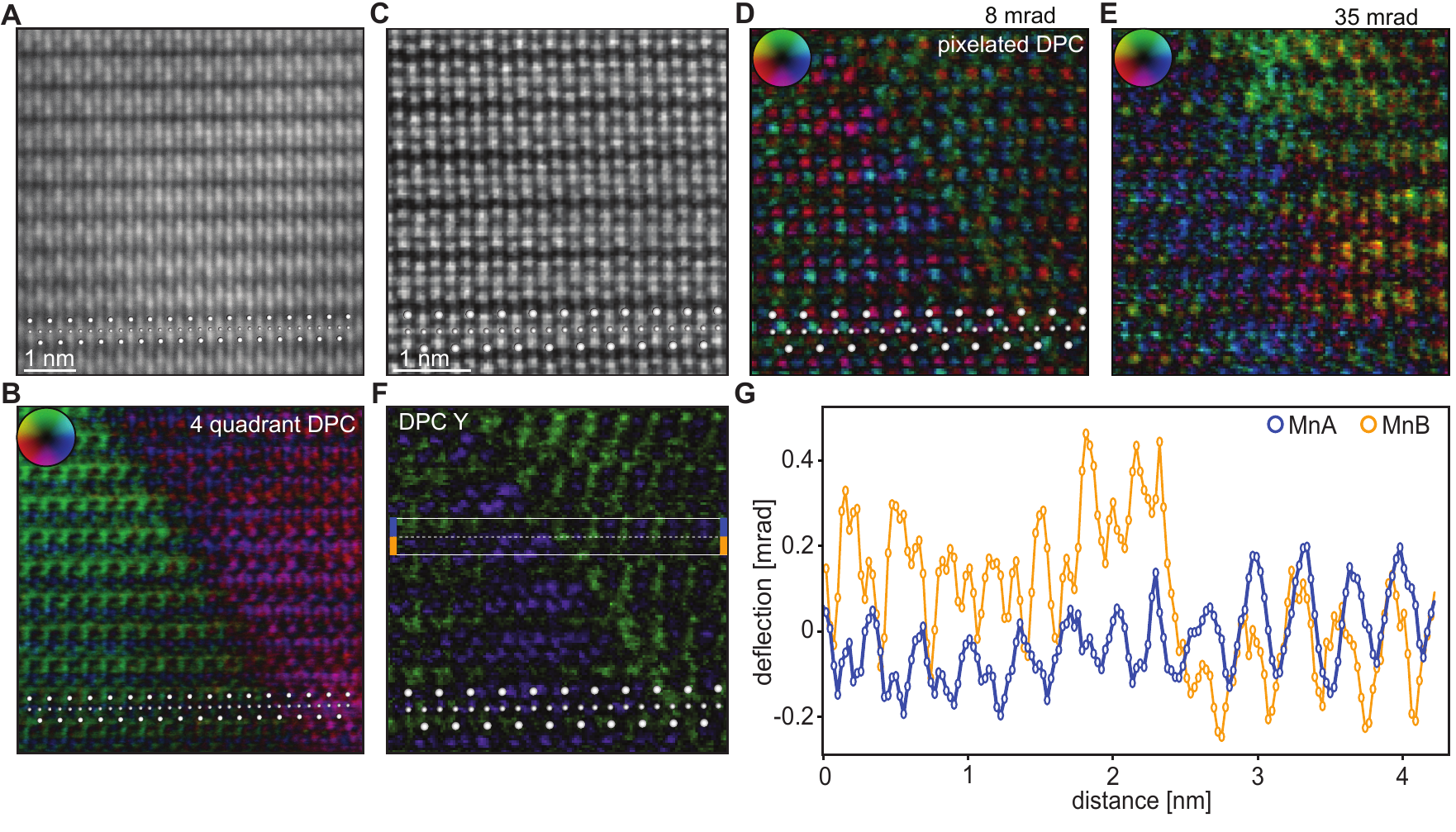}
\vspace{-0.4cm}
\caption{{\bf{DPC-STEM measurements of atomically sharp domain walls in the unperturbed part of single-crystal CuMnAs.}} ({\bf A,B})  HAADF-STEM micrograph of a pristine CuMnAs crystal and corresponding 4-quadrant DPC-STEM image, respectively.  The latter shows a sharp contrast on opposite sides of an abrupt domain wall. ({\bf C,D})  HAADF-STEM micrograph and corresponding pixelated detector DPC-STEM image processed with  a 8 mrad ronchigram mask. The latter again shows a sharp domain contrast as in (B) on a different sample. ({\bf E}) Same as (D) with a 35 mrad mask. ({\bf F}) Up/down (DPC-Y) deflection extracted from the  image in (D). ({\bf G}) DPC-Y Deflection profiles plotted separately over the Mn$_{\rm A}$ and Mn$_{\rm B}$ sublattice.}
\label{fig4}
\end{figure*}

In the bottom plots of Fig.~\ref{fig3}B, the lateral dependence of the electron beam deflection is quantified on both sides of the anti-phase boundary with sub unit-cell resolution by separately integrating the DPC-STEM signal over the crystal sublattice Mn$_{\rm A}$ and Mn$_{\rm B}$. We observe a clear difference in the deflection by the two sublattices and the difference reverses at the anti-phase boundary.

The observation that the DPC-STEM signals on the two sides of the anti-phase boundary cannot be mapped one on the other by a simple lattice translation is consistent with the presence of the antiferromagnetic domain wall, as depicted in Fig.~\ref{fig1}C, and with the special symmetry of CuMnAs where the Mn$_{{\rm A}}$ and Mn$_{{\rm B}}$ crystal sites are not connected by a translation. The reversed deflections by the two Mn sublattices on the opposite sides of the boundary have no apparent structural explanation while they are readily consistent with the antiferromagnetic domain wall model of Fig.~\ref{fig1}C.  Regarding the amplitude of the deflection, a first comparison to the experimental data in Fig.~\ref{fig3}B can be made by considering the direct effect of the Lorentz force on the electron beam (see Fig.~\ref{fig1}C). Here the deflection of the beam is given by $e\lambda B t/h$, where $e$ is the electron charge, $\lambda$ is the de Broglie wavelength of the accelerated electron, $B$ is the component of the internal magnetic field orthogonal to the beam direction, $t$ is the thickness of the lamella used in the STEM measurement, and $h$ is the Planck's constant  \cite{McVitie2015}. From our density functional theory (DFT) calculations \cite{Blaha_2001}, we obtain a strong internal magnetic field of 6~T averaged over one spin-sublattice plane of Mn atoms, with peak values reaching 20~T. For the 100~kV accelerated electrons  and lamellae with thicknesses above 50~nm, used in our measurements, the beam deflection estimated from the Lorentz force expression is close to one mrad, which is consistent with the scale of the experimental data in Fig.~\ref{fig3}B. 

We emphasize that the above estimate represents  a simplified interpretation of the measured DPC-STEM signals. For our experimental lamella thicknesses, deflection from internal electric fields and dynamical diffraction intertwine with the Lorentz deflection and inhibit a  quantitative interpretability of the DPC-STEM data \cite{Edstrom2019} (see Supplementary text for more details). The convoluted nature of the signal is apparent when comparing the same data processed with different circular aperture masks applied to the recorded CBED patterns. The comparison is shown in the bottom left panel of Fig.~\ref{fig3}A and Fig.~\ref{fig3}C, processed with 8 and 35 mrad mask radius, respectively. 

The distinction of the DPC-STEM signals on the two sides of the anti-phase boundary, which we ascribed in Fig.~\ref{fig3}A to opposite antiferromagnetic domains, fade out when using the larger 35 mrad (Fig.~\ref{fig3}C). Since  unperturbed crystals on the sides of the anti-phase boundary would be indistinguishable without magnetism, Fig.~\ref{fig3}C indicates that with the increased mask radius the electric component dominates the DPC-STEM image.  Consistently, the local changes in the crystal at the position of the anti-phase boundary defect are highlighted by the larger mask, as seen in Fig.~\ref{fig3}C.

We qualitatively confirm the above trend in the relative strength of the magnetic and electric components by performing numerical simulations of the DPC-STEM signal. We employ a state-of-the-art Pauli multi-slice method using internal magnetic and electric fields obtained from the DFT calculations  \cite{Edstrom_2016_1, Edstrom_2016_2,Edstrom2019}. For the 8 mrad mask, the simulations in Fig.~\ref{fig3}D show a significant relative contribution from the magnetic component in large portions of the CuMnAs unit cell, while  the 35 mrad mask makes the magnetic component  less visible. Simulations shown in Fig.~\ref{fig3}E also confirm that  the smaller mask combined with lamella thicknesses above 50~nm, as used in our experiments,  are favorable for making the magnetic component of the DPC-STEM signal more prominent.
 
The atomically sharp domain walls are not necessarily connected with the anti-phase boundaries. Fig.~\ref{fig4} shows two examples recorded by two different microscopes of a domain wall formed in an unperturbed part of the single-crystal (cf. Fig.~\ref{fig1}D); for another example  see Supplementary Fig.~S1. The absence of the anti-phase defect in the explored portion of the CuMnAs single-crystal epilayer is confirmed by the HAADF-STEM measurements. The DPC-STEM images, on the other hand, show two distinct domains separated by an abrupt domain wall. The basic characteristics are analogous  to the DPC-STEM data on the opposite sides of the anti-phase boundary, seen in Fig.~\ref{fig3}. This includes the reversed deflections by the Mn$_{\rm A}$ and Mn$_{\rm B}$ sublattices on the opposite sides of the atomically sharp domain wall (Figs.~4G), and also the loss of contrast between the opposite domains when using larger masks  to obtain the DPC-STEM images (Fig.~\ref{fig4}D compared to Fig.~\ref{fig4}E). On the other hand, unlike Fig.~\ref{fig3}, a crystallographic defect at the domain wall does not emerge when processing the DPC-STEM data in Figs.~4D,E with either a small or larger ronchigram mask. This is consistent with the simultaneously acquired HAADF-STEM image in Fig.~\ref{fig4}C, and the interpretation that the antiferromagnetic domain wall seen in the DPC-STEM images in Fig.~\ref{fig4} can also form in an unperturbed part of the crystal.

The abrupt reversal of the N\'eel vector at the anti-phase boundary is an expected consequence of the defect's crystallography in CuMnAs. However, the formation of an atomically sharp domain wall in an unperturbed part of a magnetic crystal is at odds with expectations of the established model spin-Hamiltonian theory (see Supplementary text and Fig.~S2). For widths below the micromagnetic domain wall ($\approx100$~nm), the theory gives a monotonic increase of the energy  with decreasing wall width, making the atomically sharp domain wall unfavorable. 

To explore the microscopic physics beyond the remit of the semiclassical model theory, we employed fully relativistic quantum-mechanical DFT implemented within the spin-polarized Korringa-Kohn-Rostoker (KKR) package \cite{Ebert2011a} (see the Supplementary text). We explored narrow walls of widths from $\sim$10 Mn atoms down to an abrupt N\'eel vector reversal between two neighboring atoms, i.e., orders of magnitude smaller than the micromagnetic domain wall width. For these narrow domain walls, our microscopic DFT calculations reveal a non-monotonic dependence of the domain wall energy on the width (Supplementary Fig.~S3). There is a drop in the energy of the atomically sharp domain wall by one third with respect to the energy of the second narrowest domain wall. The result points towards a higher stability of an atomically abrupt domain wall than expected from the model spin-Hamiltonian theory. Consistent with experiment, our DFT calculations also suggest that the domain wall angle, as shown in Fig.~\ref{fig1}D and observed experimentally in Fig.~\ref{fig4}, gives a significantly lower energy than higher angles or a vertical [001]-orientation of the domain wall (see  Supplementary text and Fig.~S4)

We now turn to the scrutiny of potential structural artifacts in the DPC-STEM images, starting from strain variations. To address this point, we compare DPC-STEM images on CuMnAs/GaP and CuMnAs/GaAs samples. The former has a high-quality fully strained CuMnAs epilayer grown on a closely lattice-matched GaP substrate, resulting in a lateral mosaic block size in CuMnAs exceeding the $\approx 400$~nm resolution limit of the employed X-ray measurement\cite{Krizek2020}. In the latter samples with a GaAs substrate, the partially relaxed CuMnAs epilayer is of a lower crystal quality with a $\approx 30$~nm mosaic block size \cite{Krizek2020}. The lattice mismatch between GaAs and CuMnAs leads to  expected and clearly visible strain and distortion gradients on both sides of the epilayer/substrate interface, as illustrated in Supplementary Fig.~S5. The images of CuMnAs/GaP, shown in Fig. \ref{fig4}E and Supplementary Fig.~S6, are strikingly different. The DPC-STEM contrast of the CuMnAs domains separated by the atomically sharp domain walls stops abruptly at the interface, leaving the GaP substrate featureless. This rules out the strain artifact interpretation. 

Additional simulations and measurements exclude the artifacts associated with local variations in material stoichiometry, lamella thickness,  and crystal rotation.  Our CuMnAs films are grown by molecular beam epitaxy at a rate of $\sim$ 8 A/min under well calibrated conditions \cite{Krizek2020}. This means that sharp local changes in stoichiometry that would be large enough to affect the beam passing through the lamellae are unlikely to occur. We confirmed this by electron energy-loss spectroscopy (EELS) measurements shown in Supplementary Fig.~S7.  The EELS data also confirm the absence of an abrupt thickness variation across the antiferromagnetic domain wall. The exclusion of the thickness variation scenario is further underpinned by DPC-STEM simulations which also rule out artifacts due to conceivable crystal rotations, as discussed in detail in the Supplementary text and Fig.~S8.

Finally, we inspect the possibility that the DPC contrast arises from CuMnAs crystals that initially had grown independently slightly shifted, and that overlap on top of each other along the electron beam direction. Anti-phase boundaries run within the epilayer along four degenerate angles corresponding to the \{011\} planes and are the only extended lattice defects observed in our STEM measurements of the high-quality CuMnAs/GaP epilayers. A crystal overlap in the layers can, therefore, be formed only by meeting of these defects when running along different \{011\} planes, as sketched in Supplementary Fig.~S9. An example of STEM images of the crystal overlap is shown in Supplementary Fig.~S10A. 

A crystal overlap can be clearly identified by HAADF-STEM but is nearly invisible in the DPC-STEM when the images are processed with an  8 mrad mask. This is one evidence which rules out that the DPC-STEM measurements of the domain walls shown in Figs.~3 and 4 with an 8 mrad mask are arising from a crystal overlap  artifact. Consistent with the structural nature of the crystal overlap, the defect becomes more clearly apparent with a larger mask (Supplementary Fig.~S10A). We again emphasize that this is an opposite trend than what is observed for the antiferromagnetic domain contrasts in Figs.~3 and 4 (see also Supplementary Fig.~S10B). 

A complementary evidence ruling out a crystal overlap artifact is that its crystallography would generate vertical intensity gradients in the STEM images on a $\approx$5~nm scale, as a result of the different angles of the anti-phase boundaries forming the crystal overlap and as confirmed  by our numerical simulations shown in Supplementary Fig.~S11A. Such gradients are, however, absent in the experimental DPC and HAADF images of the  antiferromagnetic domains (see Fig.~\ref{fig1}E and Supplementary Figs.~S6,S11B).

To conclude, our results open the basic research front of atomically sharp magnetic domain walls. Illustrating the ultimate domain wall scaling limit in an antiferromagnet gives the field an opportunity to exploit the rich materials landscape of these abundant and diverse symmetry-type  systems \cite{Zelezny2018,Nemec2018,Gomonay2018,Smejkal2018,Duine2018,Baltz2018,Siddiqui2020}.  Making the observation in CuMnAs  has also immediate consequences for spintronic device physics and engineering. It sheds light on the recently observed quenching into high-resistive nano-fragmented domain states in analog memory devices \cite{Kaspar2019} with potential neuromorphic \cite{Kurenkov2020} and ultra-fast optical applications \cite{Kimel2019}. Combined with the earlier demonstrated efficient control of the domain wall motion by electrical currents \cite{Gomonay2016,Wadley2018,Janda2020}, it also opens the prospect of coding information in individual atomically sharp antiferromagnetic domain walls. 
\vspace{0.5cm} 

\noindent{\large\bf Acknowledgements}
\vspace{0.4cm} 

\noindent We acknowledge fruitful interactions with Josef Zweck, Ondrej Sipr, Sergiy Mankovsky, Ilja Turek, and Jan Minar, and  useful discussions on DPC with Michael Zachman who acquired the DPC-STEM image shown in Supplementary Fig. S1. This work was supported by the Ministry of Education of the Czech Republic Grants LM2015087, LNSM-LNSpin, LM2018140, Czech Science Foundation Grants No. 19-28375X, 19-18623Y, the University of Nottingham EPSRC Impact Acceleration Account grant No. EP/K503800/1, the EU FET Open RIA Grant No. 766566, the Max Planck Partner Group Grant, and the Neuron Endowment Fund Grant, National Grid Infrastructure MetaCentrum provided under the programme "Projects of Large Research, Development, and Innovations Infrastructures" (CESNET LM2015042) and Innovations project 'IT4Innovations National Supercomputing Center -- LM2015070. We also thank Diamond Light Source for the provision beamtime under proposal number MM22437. The atomically resolved STEM DPC experiments and the lamellae preparation were supported by the Center for Nanophase Materials Sciences (CNMS), which is a U.S. Department of Energy, Office of Science User Facility,  and also by the CzechNanoLab project LM2018110, CEITEC Nano Research Infrastructure. J.R. acknowledges Swedish Research Council for financial support. Multislice calculations were performed on resources provided by the Swedish National Infrastructure for Computing (SNIC) at NSC Center. 

\bigskip
\vspace{-0.4cm} 
\noindent {\em Copyright notice}: This manuscript has been authored by UT-Battelle, LLC under Contract No. DE- AC05-00OR22725 with the U.S. Department of Energy. The United States Government retains and the publisher, by accepting the article for publication, acknowledges that the United States Government retains a non-exclusive, paid-up, irrevocable, world-wide license to publish or reproduce the published form of this manuscript, or allow others to do so, for United States Government purposes. The Department of Energy will provide public access to these results of federally sponsored research in accordance with the DOE Public Access Plan (http://energy.gov/downloads/doe-public-access-plan).

\onecolumngrid
\section*{\Large{Supplementary information}}

\setcounter{figure}{0} \renewcommand{\thefigure}{S\arabic{figure}}
\noindent\underline{\bf Experimental methods}

\vspace{0.3cm}

\noindent {\bf Crystal growth.}  The description of the growth of CuMnAs thin films by molecular beam epitaxy on GaP and GaAs substrates, and the detailed characteriaztion and optimization of their properties are given in Ref.~\cite{Krizek2020}. Data shown in the main text are on optimized lattice-matched CuMnAs(50nm)/GaP films, and are complemented in this Supplementary Materials by measurements on CuMnAs(50nm)/GaAs. 

\smallskip
\noindent {\bf XMLD PEEM.} The X-ray magnetic linear dichroism photoemission electron microscopy (XMLD-PEEM) measurements were performed on beamline I06 at Diamond Light Source using linearly polarized X-rays at grazing incidence of 16$^\circ$ to the sample surface. Sensitivity to the antiferromagnetic spin axis was obtained from the asymmetry of images with the X-ray energies tuned to the Mn L$_3$ absorption edge ($2p_{3/2}$ $\rightarrow$ $3d$ transitions). Spatial contrast in the asymmetry images arises from the local variation of the angle of the spin axis with respect to the X-ray polarization vector $E$. The measurements were performed with $E$ parallel to the [100] and [110] crystal axes, which are in-plane. The XMLD spectrum has a similar shape but opposite sign for both cases \cite{Wadley2017} so that dark and light areas correspond to perpendicular and parallel spin axis with respect to $E$ for $E$ parallel to [100], and vice versa for $E$ parallel to [110]. The sample environment was cooled to $\approx100$~K which increases the XMLD contrast, but does not significantly affect the size or shape of the magnetic domain pattern.

\smallskip
\noindent{\bf Lamellae preparation.} The as-grown samples were cleaved into $5\times5$~mm chips and adjusted onto standard pin stubs by means of silver lacquer. The surface was coated with $10 - 15$~nm of carbon using Leica ACE600. The scanning transmission electron microscopy (STEM) lamellae were then fabricated in FEI Helios 660 G3 FIB/SEM instrument following the commonly used protocol, utilizing electron and ion beam deposited tungsten as a protective cap. Final polishing was done at 2~kV and 25~pA. The resulting thickness varied from $\approx 50-150$~nm in the regions of interest. The estimated thickness for the images in Fig. 3 and 4 of the main text is around 50 nm, as shown below in Supplementary Fig.~S3. 

\smallskip
\noindent{\bf STEM measurements.} The prepared lamellae were investigated by three different high resolution scanning transmission electron microscopes (TEMs). The aberration-corrected Nion UltraSTEM microscope operated at 100 kV acceleration voltage was used to acquire images presented in Figs.~3A-C and 4C-G of the main text and Supplementary Figs.~S10 and S11. The images were acquired  using semi-convergence angles of 25 mrad or 18 mrad with a probe current of $\approx58$~pA.  These images were acquired from regions of the lamellae with thickness around 50~nm. The DPC signal was recorded with a pixelated (known also as universal) detector Nion 2020 Ronchigram camera, equipped with a Hamamatsu ORCA ultra-low noise scientific CMOS sensor with a 2048$\times$2048 pixel display. The CBED patterns had an average size of an average of 80 $\times$ 80 pixels with an average pixel size of 1.2 mrad.

The Thermo Fisher Scientific (TFS) TEM Titan Themis 60-300 cubed microscope operated at 300~kV acceleration voltage was used to acquire images and EELS measurements presented in Fig.~1B,E and Fig. 4A,B of the main text and Supplementary Figs. S5-S7. The STEM images were  acquired with a convergence semi-angle of 10~mrad and probe current $\approx$30~pA. The DPC images were recorded by a TFS 4-quadrant DPC annular detector. For the DPC analysis, the TEM was aligned with camera length of 580~mm (Figs. 1, 4, S5, S6) and 460~mm (Fig.~S7), where the annular detector had the collection angle of $2-13$~mrad and $3-11$~mrad, respectively. The sharp domain wall DPC contrast was observable for all studied lamella thicknesses of $\approx 50 - 150$~nm.

The Jeol NEOARM microscope operated at 200~kV acceleration voltage was used to acquire images presented in Supplementary Fig.~S1. The imaging was done using a 28~mrad convergence semi-angle. The image was acquired from regions of the lamella with thickness around 50~nm. The DPC signal was recorded in a direct electron pixelated detector, binned to 132$\times$ 132 pixels, and using a pixel size of 1.5 mrad. 

For measurements on the Nion and Jeol microscopes, the lamellae were heated for 10~hours at 80~$^{\circ}$C under vacuum before loading the samples in the microscope columns.

\smallskip
\noindent{\bf Processing of 4-D data acquired by the CCD detectors.} The 4-D data sets acquired by the Nion and Jeol microscopes were reconstructed via modified scripts available on the open source Nion Swift Python package {\cite{Nion}}. The DPC signals were reconstructed by analysing either a vertical or total shift of the center of mass of the ronchigrams, which were recorded by the pixelated detector for each pixel of acquired HAADF image. 8 and 35~mrad radius circular apertures were applied as masks, filtering the outer part of the ronchigrams. The 4-quadrant detector based DPC images were acquired using TFS software VELOX v.2.8 with a DPC plugin and are presented without further  post-processing.

\newpage

\noindent\underline{\bf Additional STEM image of the antiferromagnetic domain wall}

\begin{figure}[h!]
\begin{center}
\hspace*{-0cm}\epsfig{width=.7\columnwidth,angle=0,file=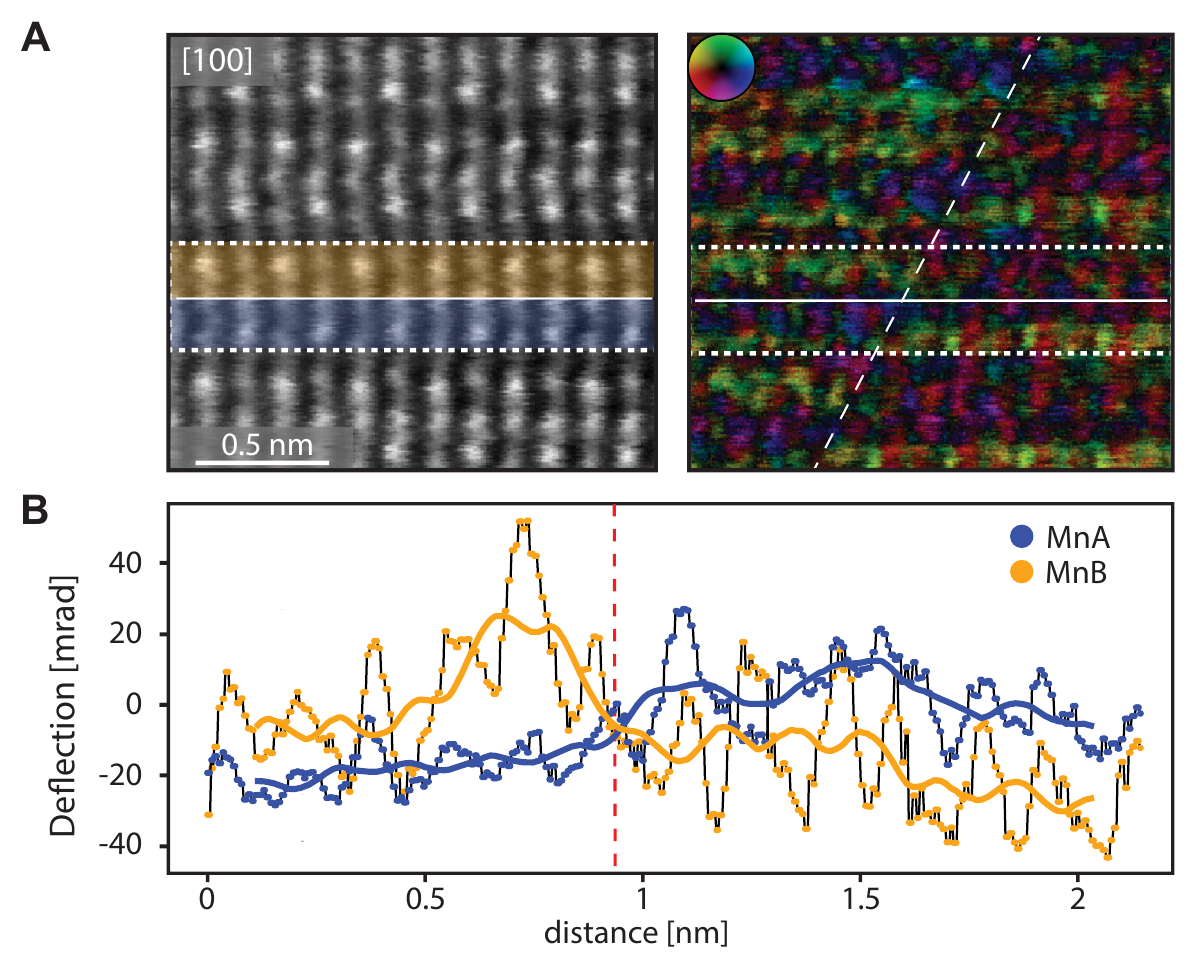}
\caption{{\bf Additional image of the antiferromagnetic domain wall} ({\bf A}) Left: HAADF-STEM image of an unperturbed part of the CuMnAs single-crystal. Right: Corresponding DPC-STEM measurement of the sharp domain wall  highlighted by a tilted white dashed-line. ({\bf B}) The vertical line profiles of total deflection (containing all directions of the ronchigram shift) over the Mn$_{\rm A}$ (blue) and Mn$_{\rm B}$ (yellow) sublattice, show an analogous trend, as in Figs.~3B and Fig.~4G in the main text. We note that compared to the main text where the signal was averaged with the Cu atoms positions excluded, here the signal includes the copper atoms. This does not affect swapping of the signal characteristics over the sharp domain boundary. The thicker lines correspond to a moving average of the data.}
\end{center}
\label{figS1}
\end{figure}

\bigskip

\noindent\underline{\bf Atomistic semiclassical Heisenberg model simulations}

\begin{figure}[h!]
\centering
\hspace*{-0cm}\epsfig{width=.5\columnwidth,angle=0,file=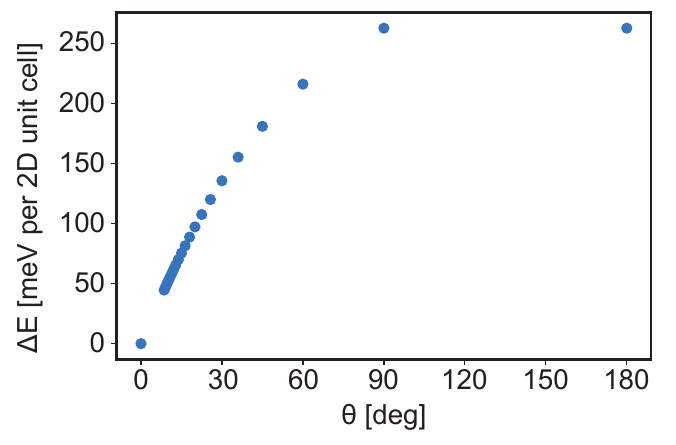}
\caption{Energy of the domain wall in CuMnAs, with respect to the energy of a uniform single-domain state, as a function of its width, parametrized  by the reorientation angle  between adjacent magnetic layers inside the domain wall. Calculations are done using the semiclassical Heisenberg spin model. The energy is plotted per 2D unit cell in the plane of the domain wall.}
\label{figS2}
\end{figure}
\smallskip
We use the atomistic spin dynamics code SPIRIT \cite{Muller2019} to simulate narrow domain walls of varying width within the semiclassical Heisenberg model. For these simulations we consider a system composed of $20\times1\times1$ unit cells with periodic boundary conditions along the [010] and [001] directions. The domain wall width is parametrized by the reorientation angle $\theta$ between adjacent magnetic layers inside the domain wall. We use exchange parameters of bulk CuMnAs calculated via the magnetic-force theorem \cite{Liechtenstein1987} using the Korringa, Kohn, Rostoker scheme in the implementation of the spin polarized relativistic KKR (SPRKKR) package \cite{Ebert2011}. We include a uniaxial anisotropy along the [001] direction of  0.1~meV per Mn,  forcing the moments to stay in the (001)-plane, and a cubic anisotropy of 0.01~meV per Mn to simulate the in-plane anisotropy.  Since we study the domain walls in the narrow regime where the energy is exchange dominated, the magnitudes of the anisotropy terms do not significantly influence the result.  

In Fig.~S2 we plot the dependence of the domain wall energy on the domain wall width parametrized via $\theta$. This dependence is similar to the result for the simplest case of a 1D ferromagnetic or antiferromagnetic chain with nearest neighbor exchange interactions. We find that the energy is increasing with decreasing width and, like in the 1D chain case, we also find that the two narrowest domains have the same energy. In the Heisenberg model with a cosine dependence of the exchange energy on the angle between neighboring spins, the atomically sharp domain wall with one neighboring spin aligned with the exchange field and another one aligned against the field has the same energy as the second narrowest domain wall with the neighboring moments pointing in the orthogonal direction. 

We note that in the 1D chain model, the configuration of spins within a domain wall of a given width is such that  all  angles between adjacent magnetic layers across the domain wall are equal. In our Heisenberg simulations of CuMnAs we find that this is also quite closely satisfied. We note that for the second narrowest domain wall with a single layer of tilted spins, the choice of the tilt angle within the Heisenberg model is arbitrary since all angles give the same energy. 
%For the ab-initio calculations we choose the tilt angle of  $90^\circ$, consistent with the spin orientation in the central layer of wider domain walls.

\bigskip

\noindent\underline{\bf DFT theory of narrow antiferromagnetic domain walls}

\begin{figure}[b!]
\centering
\hspace*{-0cm}\epsfig{width=1\columnwidth,angle=0,file=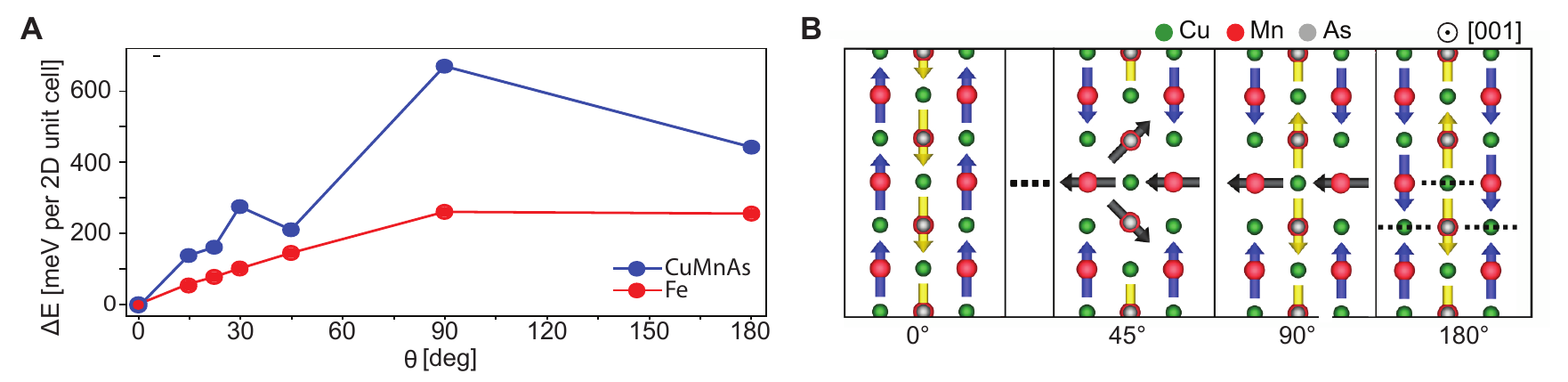}
\caption{{\bf{DFT calculations of the energy vs. width of narrow domain walls.}} ({\bf A}) Energy of the domain wall with respect to the energy of a uniform single-domain state as a function of the reorientation angle between adjacent magnetic layers inside the domain wall for antiferromagnetic CuMnAs and ferromagnetic Fe. The energy is plotted per 2D unit cell in the plane of the domain wall. ({\bf B}) Schematics of the Mn magnetic moment alignment (arrows) in CuMnAs for the first, second, and third narrowest domain wall and for the uniform single-domain antiferromagnetic state. An analogous model was considered for Fe, but only with the corresponding ferromagnetic order in the domains.} 
\label{figS3}
\end{figure}
\smallskip
\smallskip
The self-consistent-field density-functional-theory (SCF-DFT) calculations were performed in terms of the four-component Dirac electronic Hamiltonian \cite{Strange1998,Eschrig2004}, which includes relativistic effects non-perturbatively. For its Green function solution, we have resorted to the SPRKKR package \cite{Ebert2011}. Adoption of the local spin density approximation (LSDA) exchange-correlation functional by Vosko, Wilk and Nusair \cite{Vosko1980} and of the atomic sphere approximation have been previously found adequate to reproduce  CuMnAs experimental trends for ground state magnetism, N\'eel transition temperature, and charge transport \cite{Maca2017,Wagenknecht2020}, already at the linearized muffin-tin orbitals (LMTO) level \cite{Turek1997}. These approximations have also shown a good agreement with ab initio results produced by very different computational frameworks such as the plane wave basis set and the pseudopotential scheme implemented within the Vienna ab-initio simulation package (VASP) \cite{Kresse1996}, as well as the fully relativistic linear combination of atomic orbitals (LCAO) scheme by the full-potential local orbital (FPLO) package \cite{Koepernik1999}. We use experimental lattice parameters from thin film X-ray diffraction measurements \cite{Wadley2013b}
%, i.e., group number 129 with $a_{lat}$=$b_{lat}=$3.820$\AA$, $c_{lat}$=6.318 $\AA$,
and include an additional $2c$ Wyckoff position for two supplementary empty sphere sublattices, to improve convergence of the multiple scattering expansion.

For our domain wall study we examine the situation of two semi-infinite regions with opposite antiferromagnetic long range orders (LROs), which meet in the middle through a N\'eel-like domain wall.
%The magnetocrystalline anisotropy was estimated to approximately 0.1 meV per Mn by means of both discrete total energy differences for magnetic orientation along $[001]$ and $[100]$ directions and using the magnetic torque formula \cite{Staunton2006}. From this  the Mn spin magnetic moments are expected to stay within the (001)-plane. % , with origin shifted 
% by $(a_{lat}/2,a_{lat}/2)$ within the magnetic unit cell 
% and respectively shifted
% 
% arrangement at the
% corners of the two square sublattices.
We assume the $[100]$ direction as the reference orientation for the N\'eel vector, and consider a domain wall in the $(100)$-plane, within which both atomic moments of the two Mn sublattices rotate until the full recovery of LRO and with the N\'eel vector in the left region pointing along the $[100]$ direction and in the right region along the $[\bar{1}00]$ direction.
The domain wall width is then defined by the number of manganese atoms over which the above rotation takes place, and parametrized by the rotation angle $\theta$ in the above semiclassical Heisenberg model simulations. In order to efficiently describe the domain wall as an extended defect, we have resorted to a 2D tight-binding KKR calculation scheme \cite{Ujfalussy1994,Wildberger1997}, which enables a detailed computational study of a finite 2D periodic inner portion of the system (interaction zone) embedded within two semi-infinite 3D periodic (left and right) regions. This numerical framework is deployed onto a fixed interaction zone thickness of 38.2 ${\rm \AA}$ along the [100] direction, corresponding to 10 CuMnAs unit cells and verified to be adequately large for smooth matching of the embedding self-energy to the left and right semi-infinite regions, with their SCF-DFT ground states previously computed as 3D periodic and subsequently pinned. 

%The calculation setup is depicted in Fig.\ref{fig: LRO CuMnAs}, as seen down the [001] direction.

In the absence of any deviation from magnetic LRO, i.e., the same orientation throughout the interaction zone as well as within the left and right semi-infinite regions, the setup reproduces well-established bulk results for all site-diagonal quantities, such as the density of states 
%(Fig. \ref{fig: LRO DOS comparison})
and the layer-projected spin-polarized charge density; and provides in particular a reference value for the total energy.
Upon setting up the left and right regions with opposite N\'eel vector orientations, and re-converging the system with the spin magnetic moments of the interaction zone set up to complete the 180$^\circ$ rotation  along the [100] direction, we obtain a new total energy for each trial scenario. The difference with respect to the LRO case provides a definition for the energy cost of the domain wall as a function of its width, from the most abrupt scenario to a thickness spanning 11 Mn atoms along the [100] direction. 

Due to the choice of the LSDA exchange-correlation functional, as opposed to more elaborate schemes such as the transverse spin-gradient scheme \cite{Eich2013} not yet broadly available for such complex calculations, we always work with a locally collinear magnetic moment within each local frame-of-reference, in which we solve the single-site scattering problem for the effective potential around every atom. 
In particular, the spin magnetic moment given by the expectation value of the $4\times 4$ $\beta \sigma_z$ Dirac matrix is free to adjust its magnitude to minimize the Kohn-Sham total energy, but not to vary the initially prescribed direction. These directions were chosen using semiclassical Heisenberg simulations as explained above in the section on Atomistic semiclassical Heisenberg model simulations.

When using this input for lower-level electronic structure calculations of total energy within the general scheme outlined above, results have been verified for a k-point convergence within the 3D/2D Brillouin zones respectively of the semi-infinite left and right regions, and of the interaction zone in the middle. The expansion in spherical harmonics has been evaluated up to $\Lambda_{max}=3$, i.e., one level beyond the highest occupied $d$ orbitals for Cu, Mn, and As elements in the atomic limit; and recovering comparable trends also when reaching larger angular momentum truncation up to $\Lambda_{max}=5$ at a frozen potential approximation level.
Basic tests of choosing a 40\% larger interaction zone, up to 53.48 ${\rm \AA}$, have been used to verify adequate size for the SCF-DFT calculation.

% I removed the "our antiferromagnet" because it does not belong to us
Our results are summarized in Fig.~S3. For comparison, we performed analogous calculations for the antiferromagnetic CuMnAs and for ferromagnetic bcc Fe as a reference. For Fe, we obtain a result  which is still consistent with the expectation from the semiclassical Heisenberg model: Since far below the micromagnetic domain wall width, the energy of the wall increases with decreasing wall width. For the last two points, the KKR-DFT energies are nearly identical.  In contrast, the quantum mechanical KKR-DFT calculations in CuMnAs give a non-monotonous dependence of the energy on the domain wall width and a drop in the energy of the atomically sharp domain wall by one third of the highest energy corresponding to the second narrowest domain wall. This highlights the qualitative departure of the physics of narrow domain walls in CuMnAs antiferromagnet from the common understanding of magnetic textures based on the semiclassical spin models.

\begin{figure}[h!]
\centering
\hspace*{-0cm}\epsfig{width=1\columnwidth,angle=0,file=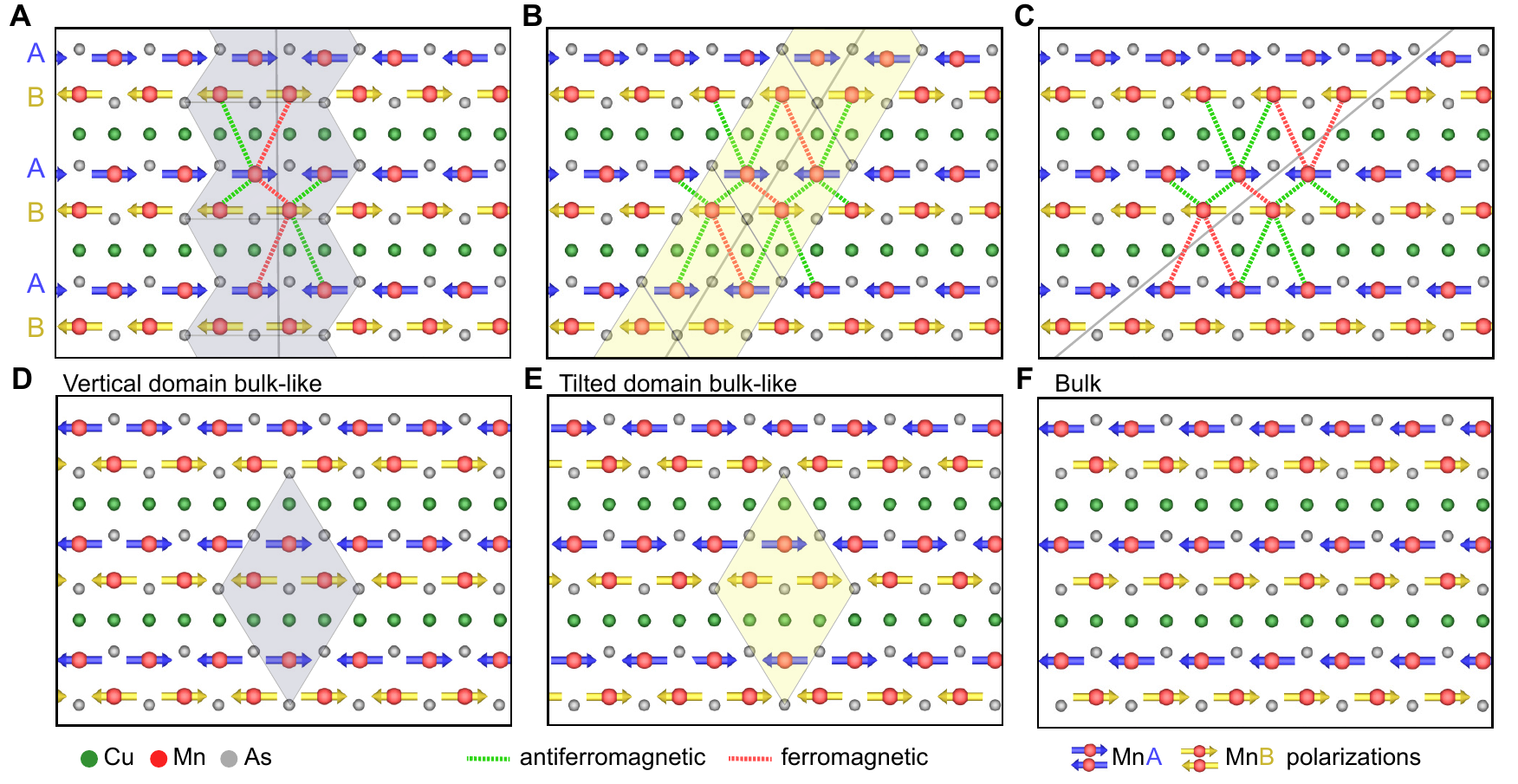}
\caption{{\bf{DFT calculations of energies corresponding to different domain wall angles}} ({\bf A,B,C}) Schematics of the vertical and tilted domain walls. Frustrated (ferromagnetic) bonds across the domain wall are highlighted by red dotted lines while unfrustrated (antiferromagnetic) bonds adjacent to the domain wall are highlighted by green dotted lines. The region of the domain wall in (A) and (B) is highlighted in grey and yellow, respectively. ({\bf D,E}) Hypothetical bulk structures with spin configurations, as highlighted by the rhombus containing only near-neighbors, corresponding to domain walls in (A,B), respectively. The hypothetical bulk structures are constructed by periodically repeating the highlighted domain wall regions in (A,B). ({\bf F}) Ground state spin configuration of bulk antiferromagnetic CuMnAs.
%NOTE: I replace rectangle for rhombus
} 

\label{figS4}
\end{figure}

The above computational KKR-DFT method is applicable to a vertical [001]-orientation of the domain wall. To estimate the relative energy differences between vertical and tilted atomically sharp domain walls we first examine the number of (ferromagnetic) bonds across the domain wall aligned against the corresponding (antiferromagetic) bulk exchange field. These frustrated bonds are highlighted by red dotted lines in the schematic plots in Fig.~S4.  In the  vertical domain wall and the domain wall at an angle observed in experiment, shown in Figs.~S4A,B, the number of frustrated bonds is the same, while the number increases for larger domain wall angles, as illustrated in Fig.~S4C. This makes domain walls with larger tilts starting from the one in Fig.~S4C unfavorable. To estimate the relative difference in energy between the domain walls in Figs.~S4A,B, we notice the different alignments of spins within the near-neighbor rectangle  which determines the magnetic configuration of the bulk. These are shown in Figs.~S4D,E in which we constructed hypothetical bulk states with the spin configurations corresponding to domain walls in Figs.~S4A,B. The hypothetical bulk structures in Figs.~S4D,E are constructed by periodically repeating  the highlighted domain wall regions in Figs.~S4A,B. We then performed DFT total energy calculations for these bulk states and obtained that the one corresponding to the tilted domain wall in Fig.~S4B, i.e. to the angle observed in experiment, has an energy which is about 20\% lower than the bulk state with the spin configuration corresponding to the vertical domain wall in Fig.~S4A. These estimates of the domain wall energies indicate that the most stable domain wall angle is the one depicted in Fig.~S4B, consistent with experiment.

\newpage

\noindent\underline{\bf Overview STEM images of strained CuMnAs/GaAs and CuMnAs/GaP interfaces}

\begin{figure}[h!]
\centering
\hspace*{-0cm}\epsfig{width=1\columnwidth,angle=0,file=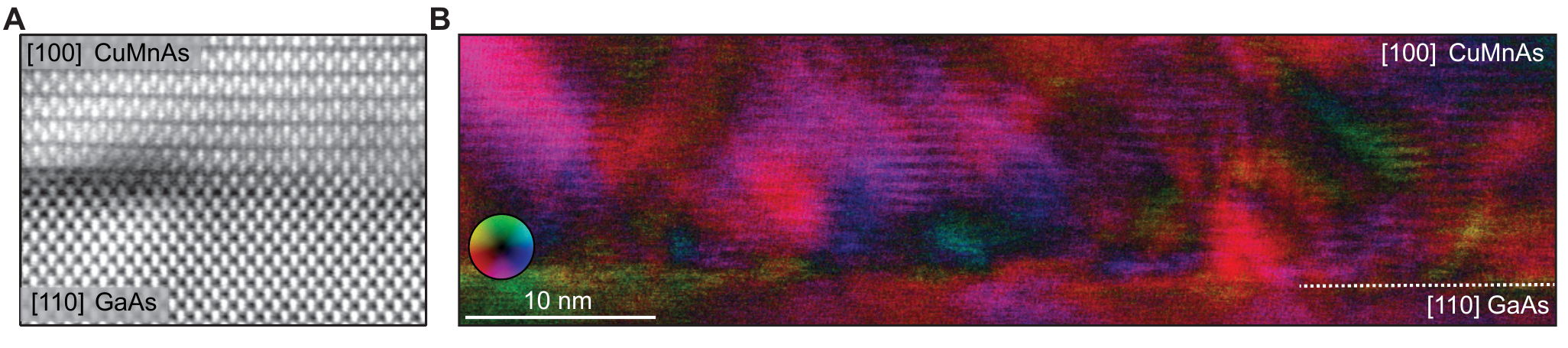}
\caption{{\bf Strain effects in CuMnAs/GaAs.}  ({\bf A}) HAADF-STEM image of the interface between 50 nm thick CuMnAs layer and a GaAs substrate \cite{Krizek2020}. Distortions attributed to the presence of misfit dislocations and strain fields are apparent in both the epilayer and the substrate. ({\bf B}) DPC-STEM image of the same layer, showing gradual changes in contrast on both sides of the epilayer/substrate interface due to the presence of strain fields.}
% No need to keep mentioning the microscope, since we have already stated that in the previous STEM methods section
\label{figS5}
\end{figure}

\begin{figure}[h!]
\centering
\hspace*{-0cm}\epsfig{width=.5\columnwidth,angle=0,file=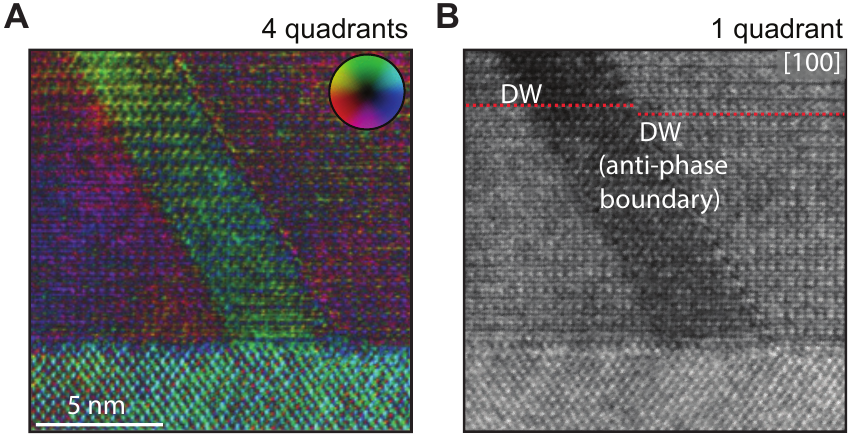}
\caption{{\bf Lattice matched CuMnAs/GaP interface.} ({\bf A}) DPC-STEM image acquired by a 4-quadrant detector of 50~nm thick CuMnAs on GaP. ({\bf B}) Same as (A) for the signal from one quadrant of the detector. The information from one quadrant makes it easier to follow the atomic lines, revealing that the right domain wall is at an anti-phase boundary while the left wall is in an unperturbed part of the crystal.}
% No need to keep mentioning the microscope, since we have already stated that in the previous STEM methods section
\label{figS6}
\end{figure}

\newpage

\noindent\underline{\bf Stoichiometry, thickness, and crystal rotation measurements and simulations}

\bigskip

\noindent {\bf Electron energy loss spectroscopy measurements.}

\begin{figure}[h!]
\centering
\vspace*{0cm}
\hspace*{-0cm}\epsfig{width=1\columnwidth,angle=0,file=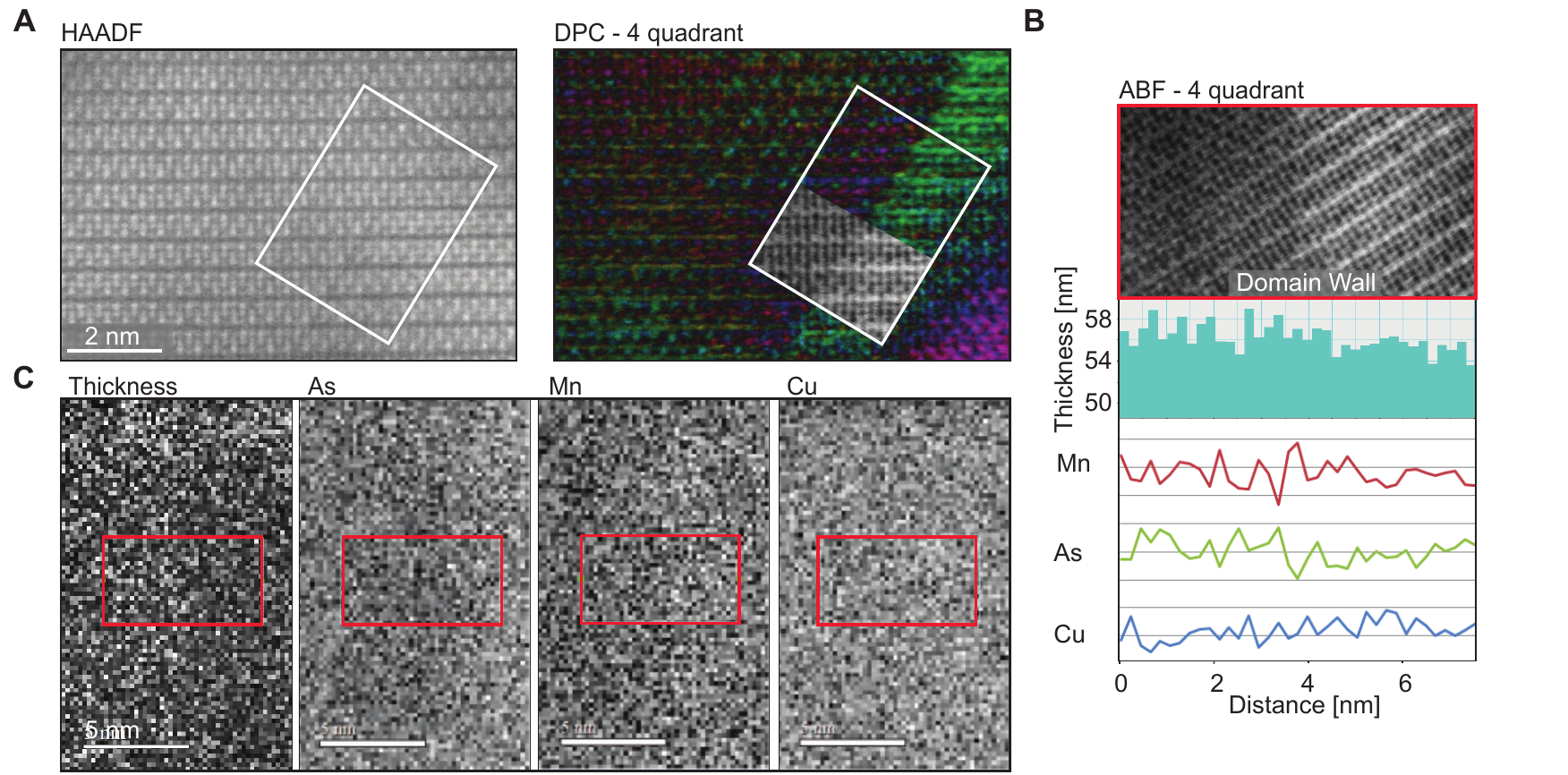}

\caption{{\bf Stoichiometry and thickness measurements.} ({\bf A}) HAADF-STEM image of a pristine CuMnAs crystal is shown in the left panel. Corresponding DPC-STEM measurement showing a sharp domain wall is presented in the right panel. The area highlighted by the white rectangle contains an inset with an image formed by summing the signals from all segments of the 4-quadrant DPC detector, i.e., the DPC detector is used as an annular bright field (ABF) detector which highlights the atomic position of the domain wall. Electron energy-loss spectroscopy (EELS) measurement were also performed in this area (as shown in (B)). ({\bf B}) Horizontal line profile of absolute thickness of the sample across the region (highlighted by red rectangle) and perpendicular to the domain wall. The thickness is calculated from a low-loss EELS spectrum image via the Log-ratio method \cite{Malis1988} and it shows no abrupt change over the domain wall. The red, green and blue intensity profiles of Mn-M$_{2,3}$, As-M$_{4,5}$ and Cu-M$_{2,3}$ edges extracted from the acquired EELS spectral image and used for profile measurements in the lower panel. The acquired element distributions confirm homogeneous distribution of Mn, As, and Cu across the domain wall. ({\bf C}) The full spatial intensity distribution of values obtained from the EELS spectrum image and used for profiles in (B).}
% No need to keep mentioning the microscope, since we have already stated that in the previous STEM methods section
\label{figS7}
\end{figure}

\newpage

\noindent {\bf Multislice simulations of structural effects.} To simulate the effects of thickness, defocus, and tilt on HAADF and DPC-STEM images of CuMnAs at atomic resolution, we have employed the DrProbe package\cite{barthel2018dr}, which implements conventional multi-slice simulations combined with the frozen phonon method. A supercell of lateral dimensions of approximately $5\times5$~nm$^2$ has been constructed (consisting of $8\times13$ unit cells with dimensions $a=0.6318$~nm and $c=0.3820$~nm). In total, 50 different frozen phonon configurations have been generated using an Einstein model and an approximate Debye-Waller factor of 0.01~nm$^{2}$. Convergence semi-angle of 25~mrad and acceleration voltage of 100~kV have been assumed, following the experimental parameters using the Nion UltraSTEM. Aberration coefficients were all set to zero, assuming a high-quality aberration correction. Only the defocus parameter has been varied in the first part of the simulations to estimate the most likely position of the focal plane in experiments. HAADF and DPC images were calculated over a single unit cell sampled by $32\times19$ beam positions, corresponding to an approximate distance of 0.02~nm between scan points. Sample thicknesses of of up to $\approx 40$~nm have been considered. In the HAADF calculations we have assumed an inner and outer collection semi-angles of 100 and 250~mrad, respectively. In the DPC simulations, the center of mass of the diffraction pattern has been calculated for a scattering angle cut-off of 8~mrad, as in the experiments shown in the main text for the smaller ronchigram mask.

\begin{figure}[h!]
\centering
\hspace*{-0cm}\epsfig{width=1\columnwidth,angle=0,file=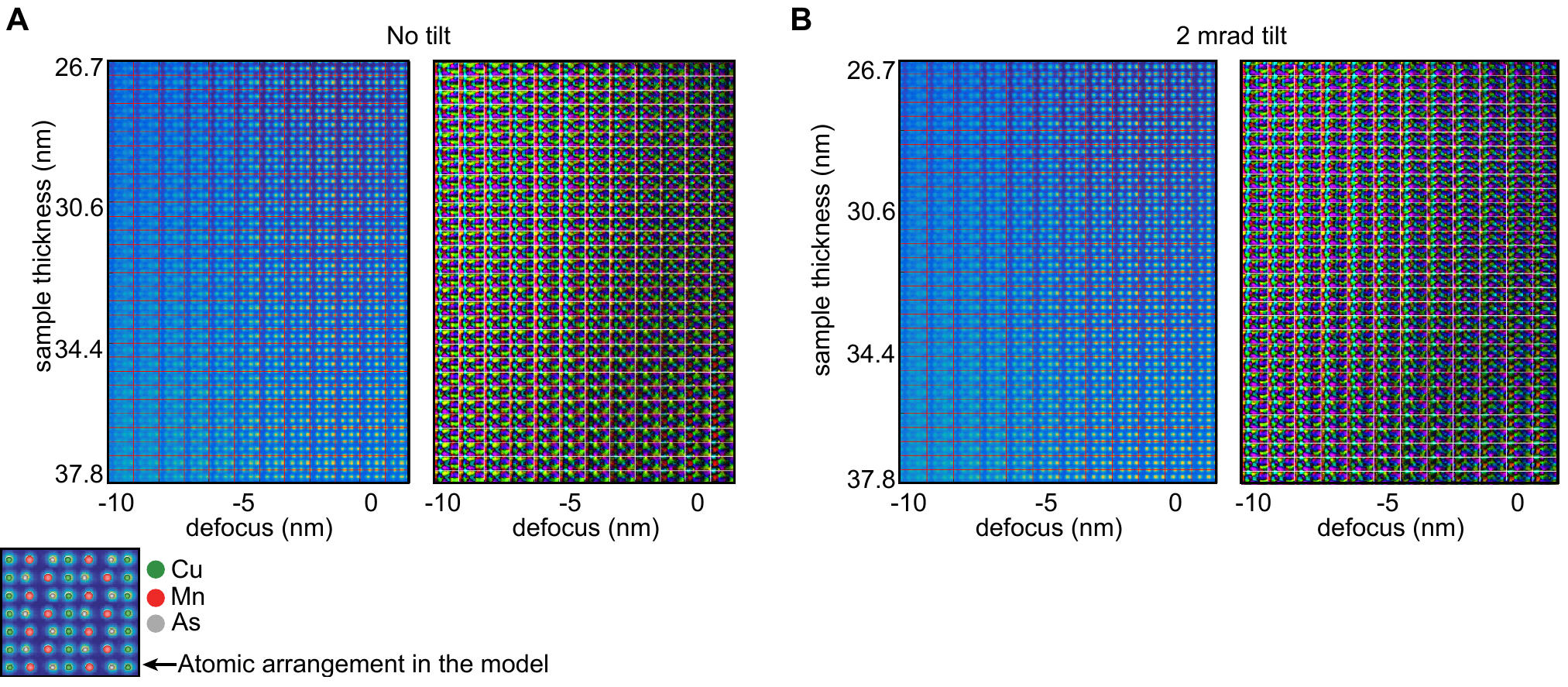}
\caption{{\bf Frozen phonon multislice calculations: thickness, defocus and tilt.} ({\bf A}) Frozen phonon multislice calculations of HAADF (left) and DPC-STEM (right) images of CuMnAs at 100~kV and 25~mrad convergence semi-angle and an electron beam exactly parallel to the zone axis. Red grid lines in the HAADF and white grid lines in the DPC image delineate regions of individual unit cells. ({\bf A}) The same as in A but with incoming beam tilted by 2~mrad from the zone axis.}
\label{figS8}
\end{figure}

In the first step, we have calculated thickness-defocus tableaux, shown in Fig.~S8A. From the simulations one can conclude: i) The defocus parameter, conventionally set in experiment to maximize the contrast in HAADF, must have been set to a value close to zero, i.e., with focal plane close to the entrance plane of the sample surface. ii) The individual images vary relatively slowly as a function of both thickness and defocus parameter. Therefore, any hypothetical thickness step (up to 3~nm in size) at the domain boundary cannot explain the large qualitative difference observed in the measured DPC images between the two domains.

Another considered structural effect is the sample tilt. We have addressed a question, whether an eventual relative crystal tilt of the two sides separated by the sharp domain wall  could potentially explain the large differences seen in the measured DPC images. For that purpose, we have repeated the simulations discussed in the previous paragraph, however this time with an incoming electron beam tilted by approximately 2~mrad from the zone axis. 
%An arbitrary tilt of (1.2~mrad, 1.6~mrad) has been randomly chosen to qualitatively check the effect of such tilt on HAADF and DPC images, respectively. 
The results are summarized in Fig.~S8B. The HAADF images are practically indistinguishable from the calculations with beam parallel to the zone axis. This matches well with the expected robustness of HAADF images with respect to small tilts. A qualitative explanation is based on the electron channeling phenomenon which overcomes the small tilt when the focused electron beam enters the sample in a close neighborhood of an atomic column. The simulated DPC image with a tilt also shows only a minor difference from the simulation where the beam was set parallel to the zone axis. The differences in simulations are substantially smaller than the qualitative changes of the experimental domain wall DPC images. Therefore we conclude that a hypothetical small relative crystal tilt also cannot explain the larger differences in DPC images observed in experiment. 

\noindent\underline{\bf Crystal overlap measurements and simulations}

%Finally, we consider crystal overlaps originating from the anti-phase boundary defects in the CuMnAs film. The crystallographic properties of CuMnAs, which dictate the possible orientations of the anti-phase boundary planes and the corresponding overlap morphology, are described  in Fig. \ref{figS9}. A comparison of STEM measurements of the crystal overlap and an unperturbed single crystal parts of the CuMnAs epilayer are shown in Fig.~S10.

\begin{figure}[h!]
\vspace*{0cm}
\centering
\hspace*{-0cm}\epsfig{width=.7\columnwidth,angle=0,file=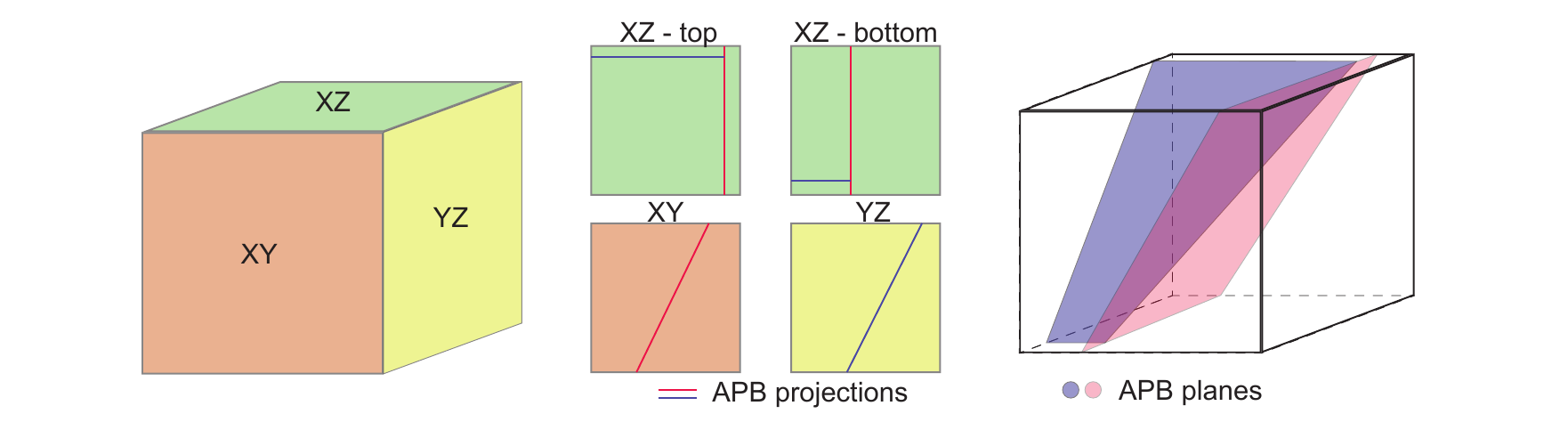}

\caption{{\bf Crystal overlap schematics.} Illustration of the formation of possible crystal overlaps in tetragonal CuMnAs thin films. Here two anti-phase boundary defects meet within the layer and form a gradual overlap, i.e., would project as vertical gradients in the HAADF-STEM images, observable from both XY [100] and YZ [010] projections of the crystal.}
\label{figS9}
\end{figure}
\begin{figure}[h!]
\vspace*{0cm}
\centering
\hspace*{-0cm}\epsfig{width=.6\columnwidth,angle=0,file=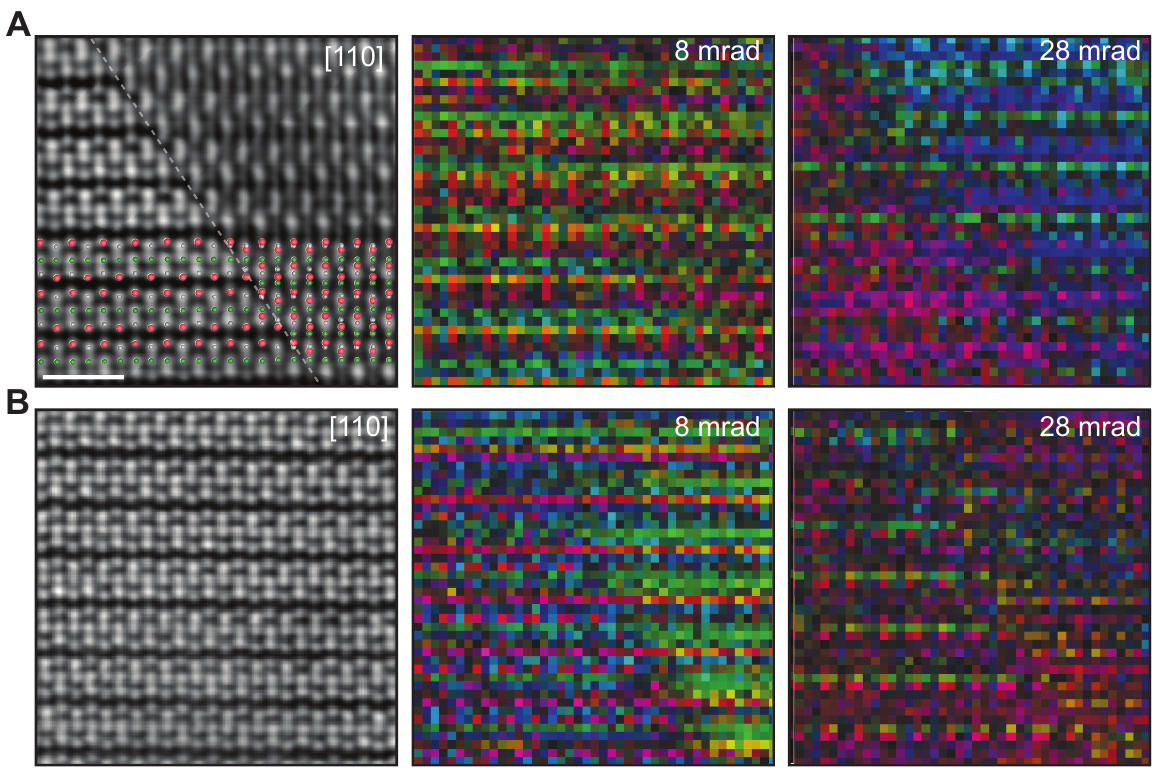}

\caption{{\bf STEM measurement of the crystal overlap.} ({\bf A}) HAADF-STEM image (left panel) showing an interface between a pristine CuMnAs crystal and a projection of an overlap region. The overlap is due to anti-phase boundaries formed along different \{011\} planes, as  illustrated in Fig.~S. The two corresponding DPC-STEM images highlight that the DPC contrast due to the overlap is partially suppressed when processing the data with  a smaller radius ronchigram mask. ({\bf B}) Same as (A) but measured in a structurally unperturbed part of the crystal containing the atomically sharp domain wall.  The DPC-STEM image shows an opposite trend to (A) with a stronger contrast appearing for the smaller mask. This is consistent with the presence of the antiferromagnetic domain wall, shown in Fig.~1D of the main text.}
\label{figS10}
\end{figure}
\begin{figure}[h!]
\centering
\vspace*{-1cm}
\hspace*{-0cm}\epsfig{width=.9\columnwidth,angle=0,file=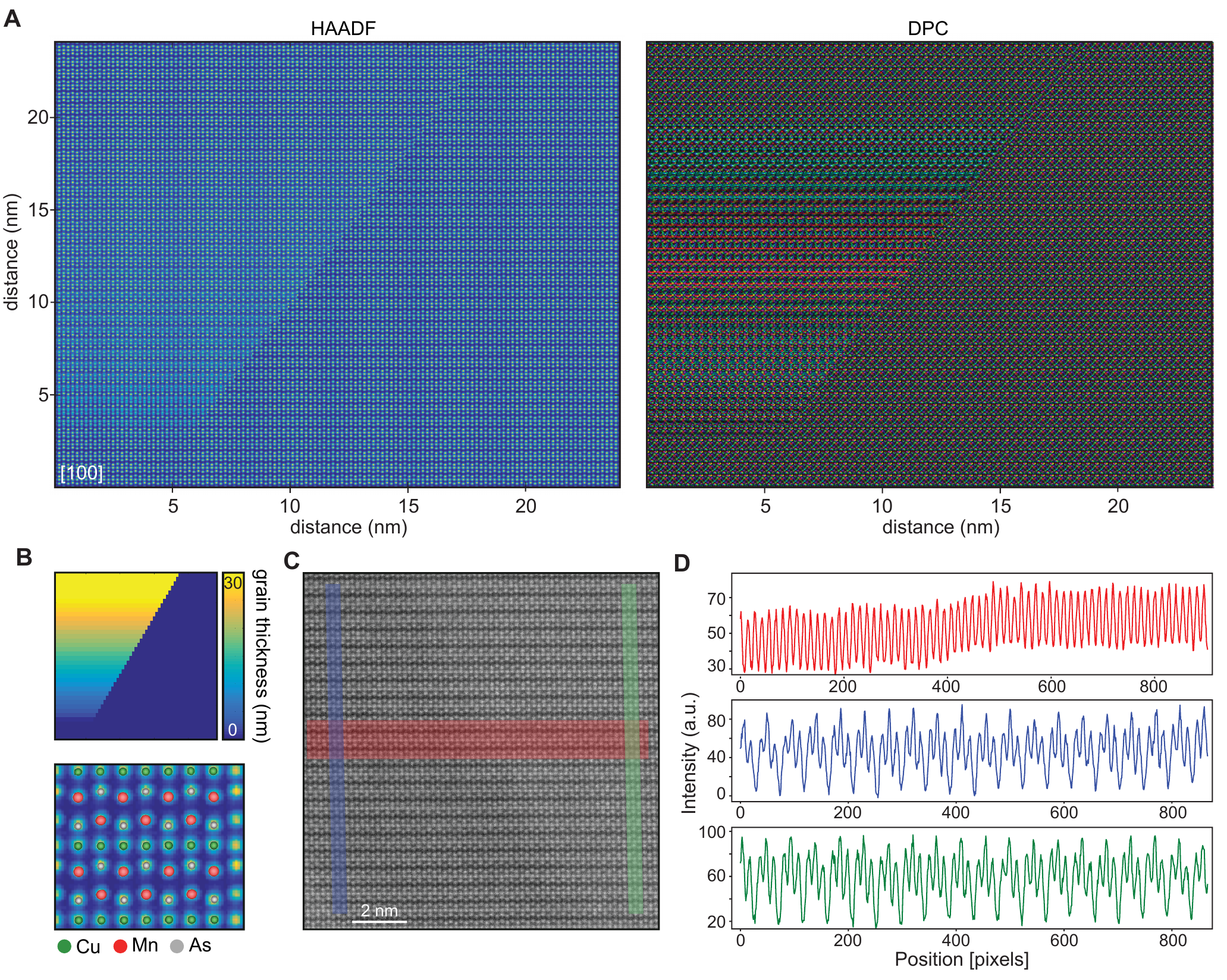}

\caption{{\bf Crystal overlap simulation compared with measurements of an unperturbed portion of the crystal featuring the sharp domain wall.} ({\bf A}) Simulations  of HAADF and DPC-STEM images  of a crystal overlap defect, showing contrast modulation in the vertical direction due to the gradually varying relative thickness between the two overlapping crystals. ({\bf B}) The morphology of the crystal overlap defect considered in the simulations, respecting the crystallographic properties of the anti-phase boundaries (see Fig.~S9). ({\bf C})  Larger field of view of the experimental HAADF-STEM image shown in Fig.~4C of the main text, corresponding to the region with the sharp domain wall seen in the DPC-STEM image in Figs.~4D-F of the main text. ({\bf D}) Vertical and horizontal intensity line profiles extracted from (C). The vertical profile directions run from top to bottom. The vertical intensity line profiles do not show any gradient. This is in contrast to the simulated STEM image of the overlap defect. The horizontal profile in (C) shows a small step in the intensity. Similarly to the DPC contrast, we found no indication that this slight change in contrast is related to common structural origins of contrast in HAADF-STEM images like, for example, the redistribution and/or change of atomic species, stoichiometry, thickness, rotation, or crystal overlap. Therefore, we surmise, that it is related to an asymmetry which, instead of a crystallography origin, is caused by the presence of the opposite antiferromagnetic domains separated by the sharp magnetic domain wall.}%
\label{figS11}
\end{figure}

\noindent{\bf Pauli multislice simulations.} Pauli multislice STEM-DPC simulations described in the main text have been performed on a grid of 32 x 16 beam positions, considering a convergent probe with convergence semi-angle of 25 mrad and acceleration voltage of 100 kV. A static lattice model has been used here. Unit cell was sampled on a grid of 128 x 64 pixels and 32 slices were generated per unit cell along the beam direction. The supercell for STEM simulations consisted of 16 x 24 unit cells in lateral directions. Magnetic component of the DPC signal was extracted as a difference of center of mass vectors extracted from two simulations with mutually reversed magnetizations.

Atomic scale DPC images are directly interpretable only at extremely low sample thicknesses and for sufficiently large collection angles\cite{Muller2017measurement}. In these conditions, the magnetic component of the DPC signal starts as an expected weak fraction of the total DPC signal. This is due to weaker interaction of electrons with magnetic fields when compared to the Coulomb potentials\cite{edstrom2019quantum}. By reducing the collection angle and increasing sample thickness, the DPC images quickly lose their quantitative and even qualitative interpretability due to pronounced dynamical diffraction effects \cite{Muller2017measurement,Burger2020influence}. Yet, perhaps paradoxically at the first sight, this complexity of dynamical diffraction effects can be of advantage for magnetic imaging. Fractionally small initial differences of DPC signals due to opposite magnetic moments in two antiferromagnetic domains will get gradually enhanced by dynamical diffraction effects, as the sample thickness increases, leading to an enhanced magnetic signal fraction (see Fig. 3E in the main text). The loss of interpretability of the spatial distribution of DPC signal can be thus traded for a higher sensitivity to the presence and qualitative changes of magnetism.

\end{document}